\documentclass[aps,preprintnumbers,nofootinbib,floatfix]{revtex4}
\usepackage{enumitem}  
\usepackage{amssymb,bm,graphicx}
\usepackage{amsmath,bbold,ulem}
\usepackage{dsfont}
\usepackage{slashed}
\usepackage[dvipsnames]{xcolor}
\usepackage[utf8]{inputenc}

\newcommand{\be}{\begin{equation}}
\newcommand{\ee}{\end{equation}}
\newcommand{\ba}{\begin{eqnarray}}
\newcommand{\ea}{\end{eqnarray}}
\newcommand{\bsub}{\begin{subequations}}
\newcommand{\esub}{\end{subequations}}
\newcommand{\la}{\langle}
\newcommand{\ra}{\rangle}

\newcommand{\blue}[1]{{\color{blue} #1}}
\newcommand{\red}[1]{{\color{red} #1}}

\allowdisplaybreaks
\begin{document}

\title{Parton model description of quark and antiquark correlators  and TMDs}

\author{
F.~Aslan$^1$, 
S.~Bastami$^1$, 
P.~Schweitzer$^1$}
  \affiliation{
  $^1$ Department of Physics, University of Connecticut, 
  Storrs, CT 06269, U.S.A.}

\begin{abstract}
Model studies play an important role for 
the understanding and elucidation of the 
nonperturbative properties of transverse
momentum dependent parton distribution
functions (TMDs).
The parton model is often a helpful
framework and starting point for first 
explorations of TMD properties and the
description of deep-inelastic processes
in which TMDs can be accessed. 
Based on a systematic exploration of the 
parton model concept, we reconcile the 
claims in literature that there are 2 
independent structures in the quark 
correlator in the parton model vs the 
claim that there are 3, and explain the 
underlying assumptions leading to the 
different conclusions.
We also systematically explore the
antiquark correlator and, to the best
of our knowledge, for the first time derive 
the model expressions for all T-even 
leading and subleading antiquark TMDs.
We demonstrate the consistency of the framework
which can be generalized in future studies for 
more sophisticated TMD modelling.

\end{abstract}

\maketitle

\section{Introduction}
\label{Sec-1:introduction}

Feynman's intuitive parton model concept
\cite{Feynman:1969ej,Feynman:1973xc} played an 
important role in establishing QCD as the theory 
of strong interactions. In many situations, the 
parton model can be considered a ``zeroth order
approximation'' to QCD \cite{Ellis:1978ty,Collins-book}. 
As such it constitutes a valuable starting point 
for explorations. This was also the case for TMDs. 
Based on a rigorous TMD factorization and evolution 
framework 
\cite{Collins-book,Echevarria:2011epo,Aybat:2011zv,
Aybat:2011ge,Collins:2014jpa,Echevarria:2016scs,
Gutierrez-Reyes:2018iod}, 
the recent past has witnessed impressive progress
in modern phenomenology of deep-inelastic scattering (DIS) processes 
\cite{Echevarria:2014xaa,Kang:2015msa,Bacchetta:2017gcc,
Bertone:2019nxa,Cammarota:2020qcw,Echevarria:2020hpy,
Kang:2020xgk,Bury:2021sue,Bacchetta:2020gko}.  
But the way to this progress was paved by, among others, 
important phenomenological work based on the 
``generalized parton model'' in works like \cite{Anselmino:1994tv,Anselmino:1999pw,
Anselmino:2005nn,Anselmino:2005sh,Anselmino:2011ch}.

A systematic exploration of the parton model concept
not for the purpose of describing DIS
processes but the nonperturbative properties of
TMDs {\it per se} was undertaken in Refs.~\cite{
Zavada:1996kp,Zavada:2001bq,Zavada:2002uz,Efremov:2004tz,
Zavada:2007ww,Efremov:2009ze,Zavada:2009ska,Efremov:2009vb,
Efremov:2010mt,Zavada:2011cv,Zavada:2013ola,Zavada:2015gaa,
Zavada:2019yom,Bastami:2020rxn,DAlesio:2009cps}, 
interestingly with conflicting~results. 
Starting from the parton model concept in one approach,
for which the name Covariant Parton Model (CMP) has been
coined, the nucleon structure is described in terms of 
two independent covariant functions
\cite{
Zavada:1996kp,Zavada:2001bq,Zavada:2002uz,Efremov:2004tz,
Zavada:2007ww,Efremov:2009ze,Zavada:2009ska,Efremov:2009vb,
Efremov:2010mt,Zavada:2011cv,Zavada:2013ola,Zavada:2015gaa,
Zavada:2019yom,Bastami:2020rxn}.
Starting from the very same parton model concept, 
the nucleon structure is described in terms of three
such independent covariant functions in another parton
model framework, namely that of Ref.~\cite{DAlesio:2009cps}. 

While technical details and especially notations may 
easily differ in independent treatments in literature,
one would expect agreement within a given approach on 
such an important point like the number of linearly 
independent structures in the quark correlator 
\cite{Bastami:2020rxn,DAlesio:2009cps} in terms
of which TMDs are defined. 
Surprisingly, there is a disagreement in literature on
this fundamental question which remains
not understood for more than a decade 
\cite{Efremov:2010mt,
Zavada:2011cv,Zavada:2013ola,Zavada:2015gaa,
Zavada:2019yom,Bastami:2020rxn,DAlesio:2009cps}.

In this work we will show that both treatments
are correct, but based on different assumptions about 
the state of polarization of a quark in the nucleon.
For that we will systematically explore the consequences
of the parton model concept for the description of the
nucleon structure and show that there are not 
``different parton model approaches'' but only one 
unifying parton model --- in which, however,
there is a choice on how to treat quark
polarization effects.
Our study will establish a bridge also to other 
works in literature, including applications of
the parton model to studies of target mass 
corrections or weak structure functions
\cite{DAlesio:2009cps,Blumlein:1996tp,
Blumlein:1996vs,Blumlein:1998nv},
early studies of quark transverse motion 
\cite{Jackson:1989ph,Roberts:1996ub} and
extensions of the parton model concept to
the free-quark ensemble model \cite{Tangerman:1994eh},
gluon polarization effects \cite{Soffer:1997zy},
or the statistical parton model approach of 
Refs.~\cite{Bourrely:2005kw,Bourrely:2005tp,Bourrely:2010ng,Bourrely:2015kla,Bourrely:2018yck}.

The structure of this work is as follows.
In Sec.~\ref{Sec-2:correlator} we will review the 
description of quark and antiquark correlators and TMDs
in models with no explicit gauge degrees of freedom
and discuss the simplifications compared to QCD.
In general, even in simpler non-gauge field 
theories care might be needed due to complications
from UV divergences, but in 
the parton model also this point is simplified 
and the correlators are UV-finite.
In Sec.~\ref{Sec-3:ampl} 
we will introduce the parton model concept and explore 
the nontrivial formal consequences arising from the 
(free) equations of motion for the quark correlator in 
massive (Sec.~\ref{Sec-4:q-correlator-massive}) and 
massless (Sec.~\ref{Sec-5:q-correlator-massless})
case following \cite{DAlesio:2009cps} where,
however, only the massless case was considered. 
In practice quark mass effects are negligible 
in DIS processes. But the distinction 
of massive vs massless partons is important when 
considering polarization effects which will turn 
out to be the key to understand and reconcile the 
conflicting results in literature. 
In Sec.~\ref{Sec-6:TMD-q}, 
we will study quark TMDs in the 
parton model with due care to the treatment
of polarization effects and reconcile the
parton model approaches of 
\cite{Bastami:2020rxn,DAlesio:2009cps}.
In Sec.~\ref{Sec-7:correlator-barq},
we will repeat the above program for the antiquark 
correlator and antiquark TMDs. 
The quark and antiquark correlators are related to
each other in a specific way in field theory. But we
will treat the quark and antiquark cases independently,
and use their field theoretic connection to demonstrate
in Sec.~\ref{Sec-8:consistency} the consistency of the
approach. 
In Sec.~\ref{Sec-9:evaluation-in-CMP} we
will solve the model and evaluate 
the parton model expressions reproducing prior
results for quark TMDs in literature 
and presenting new results for antiquark TMDs.
Finally, Sec.~\ref{Sec-11:conclusions} contains the conclusions.

\newpage
\section{Quark and antiquark correlators 
and TMDs in quark models}
\label{Sec-2:correlator}

In this section, we will review the properties of
the correlators, and the definitions of TMDs. We will 
specifically indicate the simplifications arising
in quark models by which we mean a model or
effective theory with quark and antiquark degrees of 
freedom, but without explicit gauge field degrees of 
freedom. 

\subsection{Quark and antiquark correlators}

In quark model approaches, the quark and antiquark
correlation functions of the nucleon are defined as
\bsub\label{Eq:correlators}
\ba\label{Eq:correlator-q}
	\Phi_{ij}^q(k,P,S) &=& 
	\int \frac{\mathrm{d}^4z}{(2\pi)^4}\;\mathrm{e}^{i k z}\,
    \langle N|\,\overline{\Psi}_j^{\,q}(0)\;\Psi_i^q(z)\,
    |N\rangle\,, \ \\
    \label{Eq:correlator-qbar}
    \Phi_{ij}^{\bar q}(k,P,S) &=&
    \int \dfrac{d^4 z}{(2\pi)^4} \; \mathrm{e}^{ik\cdot z}\,
    \langle N|\,\Psi_i^q(0)\;\overline{\Psi}_j^{\,q}(z)\,
    |N\rangle,
\ea\esub
where $k^\mu$ is the quark 4-momentum and $|N\ra = |P,S\ra$
denotes a covariantly normalized nucleon state with 4-momentum $P^\mu$ and
polarization $S^\mu$ with $P^2=M^2$, $S^2=-1$, 
$P\cdot S =0$. 
In QCD and in models, the correlators 
in (\ref{Eq:correlators}) are 
connected to each other as 
$\Phi_{ij}^{\bar q}(k,P,S) = - \,\Phi_{ij}^q(-k,P,S)$.
For our purposes it will, however, be
convenient not to explore this connection 
and treat the quark and antiquark cases
independently in the next sections,
until in Sec.~\ref{Sec-8:consistency}
we will come back to this connection and 
make use of it to demonstrate the 
consistency of the approach.

In QCD, the quark fields in the correlators (\ref{Eq:correlators}) 
are connected by Wilson lines which can be chosen along 
process-dependent paths dictated by factorization theorems
such that the correlators, upon integration over $k^-$ and
tracing with the relevant Dirac matrices $\Gamma$, yield
TMDs describing semi-inclusive DIS, Drell-Yan or other processes
\cite{Sivers:1989cc,Mulders:1995dh,Boer:1997nt,Collins:2002kn,Brodsky:2002rv,Belitsky:2002sm,Bomhof:2006dp,Bacchetta:2006tn,Goeke:2005hb}.
In quark models, the Wilson lines
are absent which brings simplifications
to the structure of the correlators.

One of the simplifications is that the correlators 
$\Phi_{ij}^a(k,P,S)$ with $a=q,\,\bar q$ have expansions
in terms of only 12 Lorentz-invariant amplitudes $A_i^a$
as follows 
\cite{Boer:1997nt} 
\begin{align}
    \label{Eq:correlator-decompose}
     \Phi^a(k,P,S) \; 
     &=
     MA_1^a + \slashed{P} A_2^a + \slashed{k}A_3^a 
     + \frac{i}{2M} \; [\slashed{P},\slashed{k}] \; A_4^a
     + i (k\cdot S) \gamma_5 \; A_5^a 
     + M\slashed{S} \gamma_5 \; A_6^a
     + \frac{(k\cdot S)}{M} \slashed{P} \gamma_5 \; A_7^a
     + \frac{(k\cdot S)}{M} \slashed{k} \gamma_5 \; A_8^a
 \nonumber\\
   & + \frac{[\slashed{P},\slashed{S}]}{2}\gamma_5\; A_9^a 
     + \frac{[\slashed{k},\slashed{S}]}{2} \gamma_5 \; A_{10}^a 
     + \frac{(k\cdot S)}{2M^2} [\slashed{P},\slashed{k}] \gamma_5 \; A_{11}^a
     + \frac{1}{M} \varepsilon^{\mu\nu\rho\sigma} \gamma_{\mu} P_{\nu} k_{\rho} S_{\sigma} \; A_{12}^a \,,
\end{align}
where $\varepsilon^{0123} = 1$. 
Notice that the correlators depend at most
linearly on the nucleon polarization vector
$S^\mu$. 

In QCD, besides $k$, $P$, $S$ the correlators depend 
on an additional 4-vector $n^\mu$ characterizing the
lightcone direction of the Wilson-lines ($n^\mu$ 
itself is slightly off-lightcone to regularize 
rapidity divergences) \cite{Collins-book}.
The presence of the vector $n^\mu$ in QCD 
makes more Lorentz structures possible in the
decomposition  (\ref{Eq:correlator-decompose}) 
which are described in terms of 20 additional 
amplitudes, often called $B_i^a$ amplitudes
\cite{Goeke:2005hb}, which are absent in 
quark models.

The amplitudes $A_i^a$ in (\ref{Eq:correlator-decompose}) are real 
functions of the Lorentz scalars $P\cdot k$ 
and $k^2$ \cite{Mulders:1995dh,Boer:1997nt}. 
The amplitudes $A_i^a$ for $i=2,\,3,\,6,\,7,\,8,\,12$ 
are chiral even, those for $i=1,\,4,\,5,\,9,\,10,\,11$ 
are chiral odd.
Another simplification in quark models is that the
T-odd amplitudes $A_i^a$ for $i = 4,\,5,\,12$ 
vanish \cite{Pobylitsa:2002fr}. We include them in
(\ref{Eq:correlator-decompose}) for completeness.

The correlators (\ref{Eq:correlators}) can
be used to define ``fully unintegrated'' parton
distributions which have important applications 
\cite{Watt:2003mx,Collins:2005uv,Collins:2007ph,
Rogers:2007ab}. In QCD the correlators contain 
divergences which simplifies in the parton model
as  will follow up in the next section where we
will introduce TMDs.

\subsection{Definition of TMDs}
\label{Subsec:def-TMD}

The TMDs of quarks and antiquarks are defined in terms 
of the correlators as follows. Introducing the lightcone
coordinates, $k^\mu = (k^+,k^-,\vec{k}_T)$ with
$k^\pm=\frac{1}{\sqrt{2}}(k^0\pm k^1)$ and 
analogously for other vectors, it is convenient to define
\be
     \Phi^{a[\Gamma]} \equiv 
     \Phi^{a[\Gamma]}(x,\vec{k}_T,P,S) =
     \iint dk^+dk^-\,\delta(k^+-xP^+)\,\frac12
     {\rm tr}\,\biggl[\Phi^a(k,P,S)\,\Gamma\biggr]\,.
\ee
The leading twist TMDs are defined as
\begin{subequations}
\label{Eqs:def-TMDs}
\ba
    \Phi^{a[\gamma^+]}
    &=& \zeta_V\Biggl[
    \blue{f_1^a}-\frac{\varepsilon^{jk}k_T^j S_T^k}{M}\,\red{f_{1T}^{\perp a}}\Biggr]\,,
    \label{Eq:TMD-pdfs-I}\\
    \Phi^{a[\gamma^+ \gamma_5]} &=& \zeta_A\Biggl[
    S_L\,\blue{g_1^a} + \frac{\vec{k}_T\cdot\vec{S}_T}{M}\,\blue{g_{1T}^{\perp a}}
    \Biggr]\;, 
    \label{Eq:TMD-pdfs-II}\\
    \Phi^{a[i \sigma^{j+} \gamma_5]} &=& \zeta_T\Biggl[
    S_T^j\,\blue{h_1^a}  + S_L\,\frac{k_T^j}{M}\,\blue{h_{1L}^{\perp a}} +
    \frac{\kappa^{jk}S_T^k}{M^2}\,
    \blue{h_{1T}^{\perp a}} + \frac{\varepsilon^{jk} k_T^k}{M}\,\red{h_1^{\perp a}}
    \Biggr]\;,  \hspace{15mm}
    \label{Eq:TMD-pdfs-III}
\ea
and the twist-3 quark TMDs are given by
\ba
\hspace{-5mm}    
	\Phi^{a[\mathbb{1}]}        &=&
    	\zeta_S\;\frac{M}{P^+}\biggl[
	\;\blue{e^a}
	-\frac{\varepsilon^{jk} k_T^j S_T^k}{M}\,\red{e_T^{\perp a}}\,
    	\biggr], \label{Eq:sub-TMD-pdfs-I}\\
\hspace{-5mm}    
	\Phi^{a[i \gamma^5]}       &=&
        \zeta_P\;\frac{M}{P^+}\biggl[
    	S_L \, \red{e_L^a} +\frac{\vec{k}_T \cdot \vec{S}_T}{M}\,\red{e_T^a}
    	\biggr], \label{Eq:sub-TMD-pdfs-II}\\
\hspace{-5mm}    
	\Phi^{a[\gamma^j]}       &=&
        \zeta_V\;\frac{M}{P^+}\biggl[
    	\frac{ k_T^j}{M}\blue{f^{\perp a}}\!+\varepsilon^{jk}S_T^k\red{f_T^a}
	\!+\!S_L\frac{\varepsilon^{jk} k_T^k}{M}\red{f_L^{\perp a}}
	\!-\!\frac{\kappa^{jk}\varepsilon^{kl}S_T^l}{M^2}\red{f_T^{\perp a}}\!
	\biggr], \label{Eq:sub-TMD-pdfs-III}\\
\hspace{-5mm}    
	\Phi^{a[\gamma^j \gamma^5]} &=&
    	\zeta_A\;\frac{M}{P^+}\biggl[
    	S_T^j\,\blue{g_T^a} 
	+ S_L\,\frac{ k_T^j}{M}\blue{g_L^{\perp a}} +
	\frac{\kappa^{jk}S_T^k}{M^2}
    	\,\blue{g_T^{\perp a}} 
	+\frac{\varepsilon^{jk} k_T^k}{M}\,\red{g^{\perp a}} 
	\biggr], \label{Eq:sub-TMD-pdfs-IV}\\
\hspace{-5mm}    
	\Phi^{a[i \sigma^{jk} \gamma^5]} &=&
    	\zeta_T\;\frac{M}{P^+}\biggl[
    	\frac{S_T^j  k_T^k-S_T^k  k_T^j}{M}\,\blue{h_T^{\perp a}}
    	-\varepsilon^{jk}\,\red{h^a} 
	\biggr], \label{Eq:TMD-pdfs-V} \\
\hspace{-5mm}    
	\Phi^{a[i \sigma^{+-} \gamma^5]} 
	&=& \zeta_T\;\frac{M}{P^+}\biggl[
    	S_L\,\blue{h_L^a} + \frac{\vec{k}_T\cdot\vec{S}_T}{M}\,\blue{h_T^a}
    	\biggr]. \label{Eq:TMD-pdfs-VI}
\ea
The spatial indices $j,\,k$ are transverse with respect 
to the lightcone which is chosen along 0- and 
1-directions, and we defined
$\kappa^{jk}=( k_T^jk_T^k-\frac12\delta^{jk} \vec{k}_T^2)$,
and $\varepsilon^{23}=-\varepsilon^{32}=1$ and zero else.
The T-even TMDs are highlighted in blue color and can be
computed in quark models. The T-odd TMDs highlighted in
red color require explicit gauge field degrees of freedom,
and are not studied in this work. 
The factors 
\ba
       \zeta_V = \zeta_T = \zeta_P = + 1 \quad 
       \mbox{for} \quad a = q,\,\bar q, \phantom{\bigl|} \hspace{-2mm}\nonumber\\
       \zeta_S = \zeta_A = 
       \begin{cases} +1 & \mbox{for}\quad a = q, \\
                     -1 & \mbox{for}\quad a = \bar{q}, 
                     \phantom{\displaystyle\frac11}\end{cases}
\ea\end{subequations}
reflect the different $C$-parities of the quark 
bilinear operators $\bar\Psi\Gamma\Psi$ which are even 
in the vector (V), tensor (T) and pseudo-scalar (P) cases 
$\Gamma=\gamma^\mu,\;i\sigma^{\mu\nu}\gamma_5,\, i\gamma_5$,
and odd in the scalar (S) and axial-vector (A) cases
$\Gamma=\mathds{1},\,\gamma^\mu\gamma_5$. 
The arguments of TMDs in (\ref{Eqs:def-TMDs}) 
are omitted for brevity with the understanding that
$f_1^a=f^a_1(x, k_T)$ where $k_T=|\vec{k}_T|$, etc.

We remark that in QCD factorization, TMDs 
depend on two scales commonly denoted as $\mu^2$ 
(renormalization scale) and $\zeta$ (scale
at which lightcone divergences are regulated).
In general, also in models divergences may occur
and require careful treatment
\cite{Schweitzer:2012hh,Aslan:2021new2}. 
But in the parton model, the correlators and TMDs
are finite, and we will throughout refrain from
indicating the dependence on the scales $\mu^2$,
$\zeta$ and comment on them when necessary.
Because of these simplifications in contrast 
to QCD, in the parton model TMDs and colinear 
parton distribution functions (PDFs) are simply related
as, e.g., $f_1^a(x) = \int d^2k_Tf_1^a(x,k_T)$ 
where the integration over $k_T$ is finite.

In this work, we will focus on T-even TMDs which are
expressed in terms of the $A_i^q$ amplitudes as follows
\bsub
\label{Eq:TMDs-q-amp}
\begin{eqnarray}
    \label{Eq:TMDq_f_1}
         f_1^q(x,k_T) \; &=& \; 
         2P^+ \int dk^- (A_2^q + x A_3^q) \, , \\
    \label{Eq:TMDq_f^perp_1T}
         g_1^q(x,k_T) \; &=& \; 
         2P^+ \int dk^- 
        \biggl( -A_6^q-\frac{P \cdot k-M^2x}{M^2}(A_7^q+xA_8^q)\biggr) \, , \\
    \label{Eq:TMDq_g_1T}
         g_{1T}^{\perp q}(x,k_T) \; &=& \; 
         2P^+ \int dk^-(A_7^q+xA_8^q) \, , \\
    \label{Eq:TMDq_h_1}
         h_1^q(x,k_T) \; &=& \; 
         2P^+ \int dk^- 
         \biggl( -A_9^q-xA_{10}^q+\frac{\vec{k}_T^{\,2}}{2M^2} \; A_{11}^q \biggr) 
         \, ,\\
    \label{Eq:TMDq_h^perp_1L}
         h^{\perp q}_{1L}(x,k_T) \; &=& \; 
         2P^+ \int dk^- 
         \biggl( A_{10}^q - \frac{P \cdot k - M^2x}{M^2} \; A_{11}^q\biggr) 
         \, , \\
    \label{Eq:TMDq_h^perp_1T}
         h^{\perp q}_{1T}(x,k_T) \; &=& \; 
         2P^+ \int dk^- A_{11}^q \, , \\
    \label{Eq:TMDq_e}
         e^q(x,k_T) \; &=& \; 
         2P^+ \int dk^- A_1^q \, , \\
    \label{Eq:TMDq_f^perp}
         f^{\perp q}(x,k_T) \; &=& \; 
         2P^+ \int dk^- A_3^q \, , \\
         \label{eq:appg_T}
         g_T^q(x,k_T) \; &=& \; 
         2P^+ \int dk^- 
         \biggl( -A_6^q + \frac{\vec{k}_T^{\,2}}{2M^2} \; A_8^q \biggr) \, , \\
    \label{Eq:TMDq_g^perp_L}
         g^{\perp q}_L(x,k_T) \; &=& \; 
         2P^+ \int dk^- 
         \biggl( - \frac{P \cdot k-M^2x}{M^2}\; A_8^q\biggr) \, , \\
    \label{Eq:TMDq_g^perp_T}
         g^{\perp q}_T(x,k_T) \; &=& \; 
         2P^+ \int dk^- A_8^q \, , \\
    \label{Eq:TMDq_h_L}
         h_L^q(x,k_T) \; &=& \; 
         2P^+ \int dk^- 
         \biggl( -A_9^q-\frac{P \cdot k}{M^2} \; A_{10}^q 
         + \biggl( \frac{P \cdot k-M^2x}{M^2} \biggr)^2 A_{11}^q \biggr) \, , \\
    \label{Eq:TMDq_h_T}
         h_T^q(x,k_T) \; &=& \; 
         2P^+ \int dk^- 
         \biggl( - \frac{P \cdot k-M^2x}{M^2} \; A_{11}^q\biggr) \, , \\
    \label{Eq:TMDq_h^perp_T}
         h^{\perp q}_T(x,k_T) \; &=& \; 
         2P^+ \int dk^- (-A_{10}^q) \, , 
\end{eqnarray}
\esub
where it is understood that $k^+=xP^+$ is fixed.
We remind that these expressions are valid in quark models.
In QCD also the $B_i^q$ amplitudes contribute, see for 
instance Ref.~\cite{Metz:2008ib}.
The antiquark TMDs are given by
\bsub
\label{Eq:TMDs-qbar-amp}
\begin{eqnarray}
    \label{Eq:TMDqbar_f_1}
         f_1^{\bar q}(x,k_T) \; &=& \; 
         2P^+ \int dk^- \biggl(A_2^{\bar q} + x A_3^{\bar q}\biggr) \, , \\
    \label{Eq:TMDqbar_f^perp_1T}
         g_1^{\bar q}(x,k_T) \; &=& \; 
         2P^+ \int dk^- 
         \biggl(A_6^{\bar q}
         +\frac{P \cdot k-M^2x}{M^2}
         (A_7^{\bar q}+xA_8^{\bar q})\biggr) \, , \\
    \label{Eq:TMDqbar_g_1T}
         g_{1T}^{\perp \bar q}(x,k_T) \; &=& \; 
         2P^+ \int dk^-\biggl(-\,A_7^{\bar q}-xA_8^{\bar q}\biggr) \, , \\
    \label{Eq:TMDqbar_h_1}
         h_1^{\bar q}(x,k_T) \; &=& \; 
        2P^+ \int dk^- 
         \biggl( -A_9^{\bar q}-xA_{10}^{\bar q}+\frac{\vec{k}_T^{\,2}}{2M^2} \; A_{11}^{\bar q} \biggr) \, , \\
    \label{Eq:TMDqbar_h^perp_1L}
         h^{\perp \bar q}_{1L}(x,k_T) \; &=& \; 
         2P^+ \int dk^- 
         \biggl( A_{10}^{\bar q} - \frac{P \cdot k - M^2x}{M^2} \; A_{11}^{\bar q}\biggr) \, , \\
    \label{Eq:TMDqbar_h^perp_1T}
         h^{\perp \bar q}_{1T}(x,k_T) \; &=& \; 
         2P^+ \int dk^- A_{11}^{\bar q} \, ,  \\ \label{Eq:TMDqbar_e}
         e^{\bar q}(x,k_T) \; &=& \; 
         2P^+ \int dk^- \biggl(-\,A_1^{\bar q}\biggr) \, , \\
    \label{Eq:TMDqbar_f^perp}
         f^{\perp \bar q}(x,k_T) \; &=& \; 
         2P^+ \int dk^- A_3^{\bar q} \, ,  \\   
    \label{Eq:TMDqbar_g_T}
         g_T^{\bar q}(x,k_T) \; &=& \; 
         2P^+ \int dk^- 
         \biggl( A_6^{\bar q} - \frac{\vec{k}_T^{\,2}}{2M^2} \; A_8^{\bar q} \biggr) \, ,  \\
    \label{Eq:TMDqbar_g^perp_L}
         g^{\perp \bar q}_L(x,k_T) \; &=& \; 
         2P^+ \int dk^- 
         \biggl( \frac{P \cdot k-M^2x}{M^2}\; 
         A_8^{\bar q}\biggr) \, , \\
    \label{Eq:TMDqbar_g^perp_T}
         g^{\perp \bar q}_T(x,k_T) \; &=& \; 
         2P^+ \int dk^- \biggl(-\,A_8^{\bar q}\biggr) \, ,  \\
    \label{Eq:TMDqbar_h_L}
        h_L^{\bar q}(x,k_T) \; &=& \; 
         2P^+ \int dk^- 
         \biggl( -A_9^{\bar q}-\frac{P \cdot k}{M^2} \; A_{10}^{\bar q} 
         + \biggl( \frac{P \cdot k-M^2x}{M^2} \biggr)^2 A_{11}^{\bar q} \biggr) \, ,  \\
    \label{Eq:TMDqbar_h^perp_T}
         h^{\perp \bar q}_T(x,k_T) \; &=& \; 
         2P^+ \int dk^- \biggl(-A_{10}^{\bar q}\biggr) \, , \\
    \label{Eq:TMDqbar_h_T}
         h_T^{\bar q}(x,k_T) \; &=& \; 
         2P^+ \int dk^- 
         \biggl( - \frac{P \cdot k-M^2x}{M^2}\;
         A_{11}^{\bar q}\biggr) \, .
\end{eqnarray}
\esub

\section{Constraints on amplitudes from
equations of motion in parton model}
\label{Sec-3:ampl}

In this section we discuss how the equations of motion
relate the amplitudes in the parton model. 
A similar analysis was presented in \cite{DAlesio:2009cps} 
for massless quarks. Here we keep track of mass terms
and extend the analysis to antiquarks.

\subsection{Quark case}
\label{Sec-3:ampl-q}

In the parton model the quark fields satisfy the Dirac equation
$(i\slashed{\partial}-m_q)\,\Psi^q(z)=0$. Writing out all Dirac
indices and performing an integration by parts, we obtain 
\ba
	&&
	\int \frac{\mathrm{d}^4z}{(2\pi)^4}\;\mathrm{e}^{i k z}\,
    \langle N|\,\overline{\Psi}_j^{\,q}(0)\;
    \Gamma_{jl}^{ }
    \biggl[(i\overrightarrow{\slashed{\partial}}-m_q)_{li}^{ }
    \Psi_i^q(z)\,\biggr]|N\rangle \nonumber\\
    && \quad \quad =
    \Gamma_{jl}^{ }(\slashed{k}-m_q)_{li}^{ }
    \int \frac{\mathrm{d}^4z}{(2\pi)^4}\;\mathrm{e}^{i k z}\,
    \langle N|\,\overline{\Psi}_j^{\,q}(0)\Psi_i^q(z)\
    |N\rangle \nonumber\\
    && \quad \quad =
    \Gamma_{jl}^{ }(\slashed{k}-m_q)_{li}^{ } \phi^q_{ij}(k,P,S)
     = {\rm tr} \biggl[\Gamma(\slashed{k}-m_q) \phi^q(k,P,S)
     \biggr] = 0 \, ,
\ea
where $\Gamma$ can be any Dirac matrix. It is convenient 
to define identities for the quark correlator (\ref{Eq:correlator-decompose}) as follows
\be
     {\rm Tr}^q[\Gamma] \equiv \frac14\;
     {\rm tr}\,\biggl[(\slashed{k}-m_q)\,\Phi^q(k,P,S)\Gamma\biggr] = 0\, .
\ee
Exploring these identities for
$\Gamma=\mathbb{1},\,\gamma_5,\,\gamma^\mu,\,\gamma^\mu\gamma_5,
\,i\,\sigma^{\mu\nu}\gamma_5$ yields relations among the
amplitudes in the parton model. It is instructive to show 
the derivation of these relations in detail.
Let us consider first $\Gamma = \gamma^\mu$ which yields
\ba\label{Eq:ampl-01}
      {\rm Tr}^q[\gamma^\mu] 
      &=& 
      k^\mu \biggl\{M A_1^q - m_q A_3^q + i \frac{(P\cdot k)}{M}\,A_4^q\biggr\}
      -P^\mu m_q \biggl\{A_2^q+i\,\frac{m_q}{M}\,A_4^q\biggr\}
      +\varepsilon^{\mu\alpha\beta\gamma} k_\alpha P_\beta S_\gamma\,
      \biggl\{iA_9^q-\frac{m_q}{M}\,A_{12}^q\biggr\} = 0\,. \quad
\ea
As the four-vectors $k_\mu$, $P_\mu$, $S_\mu$ and 
$\varepsilon_{\mu\rho\sigma\tau}k^\rho P^\sigma S^\tau$ are
linearly independent, each of the expressions in the curly
brackets must vanish separately. Since the $A_i$'s are real, 
in each of the curly brackets the real and the imaginary parts 
must vanish separately. In this way, we find the relations
\ba\label{Eq:ampl-02}
        A_1^q = \frac{m_q}{M}\,A_3^q, \quad
        A_2^q = 0, \quad
        A_4^q = 0, \quad
        A_9^q = 0, \quad 
        A_{12}^q = 0.
\ea
Exploring ${\rm Tr}^q[\mathbb{1}]$ yields no new information
beyond what we found in (\ref{Eq:ampl-02}). 
Considering $\Gamma = \gamma^\mu\gamma_5$ yields
\ba\label{Eq:ampl-03}
      {\rm Tr}^q[\gamma^\mu\gamma_5] 
      &=& 
      k^\mu (k\cdot S)\,\biggl\{-iA_5^q  +\frac{m_q}{M}\,A_8^q 
      + A_{10}^q -\frac{(P\cdot k)}{M^2}\,A_{11}^q \biggr\} 
      \nonumber\\
      &+& P^\mu (k\cdot S)\,\biggl\{\frac{m_q}{M} A_7^q +A_9^q
      + \frac{k^2}{M^2}\,A_{11}^q\biggr\} \nonumber\\
      &+& S^\mu\,\biggl\{ m_q M  A_6^q - (P\cdot k)A_9^q - k^2A_{10}^q\biggr\} \nonumber\\
      &=& 0 \, .
\ea
We again explore that the $A_i^q$'s
are real and $k_\mu$, $P_\mu$, $S_\mu$,
$\varepsilon_{\mu\rho\sigma\tau}k^\rho P^\sigma S^\tau$
are linearly independent such that the real and
imaginary parts in each curly bracket in 
(\ref{Eq:ampl-03}) vanish. Considering that 
$k^2 = m_q^2$ and using $A_9^q=0$ from
(\ref{Eq:ampl-02}) we obtain
\be\label{Eq:ampl-04}
      A_5^q = 0, \quad
      A_6^q = \frac{m_q}{M}\,A_{10}^q, \quad
      A_7^q = -\frac{m_q}{M}\,A_{11}^q, \quad
      A_{10}^q = \frac{(P\cdot k)}{M^2}\,A_{11}^q -\frac{m_q}{M}\,A_8^q \,.
\ee
Considering ${\rm Tr}^q[\gamma_5]$ or 
${\rm Tr}^q[i\sigma^{\mu\nu}\gamma_5]$ gives no 
new information beyond (\ref{Eq:ampl-04}).

Several comments are in order. The T-odd amplitudes 
$A_4^q$, $A_5^q$, $A_{12}^q$ vanish in the parton model 
which is expected as we deal with a quark model without 
explicit gauge field degrees of freedom \cite{Pobylitsa:2002fr}. Interestingly, in 
the parton model also the T-even amplitudes $A_2^q$ and $A_9^q$
vanish. In QCD, these amplitudes receive contributions from
"genuine twist-3" quark-gluon correlators and are non-zero. 
In the parton model, the amplitudes $A_1^q$,
$A_6^q$, $A_7^q$ are proportional to current quark masses. 
In QCD, these amplitudes contain, besides mass
terms, also contributions from quark-gluon correlators.

For our purposes it is important to notice that, after exploring the
equations of motion, all non-zero amplitudes are related in one or
another way to $A_3^q$, $A_8^q$, $A_{11}^q$, i.e.\ to one unpolarized
amplitude ($A_3^q$), one chiral even polarized amplitude ($A_8^q$),
and one chiral odd polarized amplitude ($A_{11}^q$).

\subsection{Antiquark case}
\label{Sec:ampl-qbar}

In the antiquark case, we start from the free Dirac equation
$\bar\Psi^q(z)(i\overleftarrow{\slashed{\partial}}+m_q)=0$ 
where the arrow indicates which field is differentiated. 
Proceeding similarly to the quark case, we have
\ba
    && \int\dfrac{d^4 z}{(2\pi)^4}\;\mathrm{e}^{ik\cdot z}\,
    \langle N|\,\Gamma_{li}
    \Psi_i^q(0)\;\biggl[\overline{\Psi}_j^{\,q}(z)
    (i\overleftarrow{\slashed{\partial}}+m_q)_{jl}^{ }\biggr]
    |N\rangle \nonumber\\
    && \quad \quad =  
    \Gamma_{li}
    \int\dfrac{d^4 z}{(2\pi)^4}\;\mathrm{e}^{ik\cdot z}\,
    \langle N|
    \Psi_i^q(0)\;\overline{\Psi}_j^{\,q}(z)
    |N\rangle\,
    (\slashed{k}+m_q)_{jl}^{ }\nonumber\\
    && \quad \quad =  
    \Gamma_{li}
    \phi^{\bar q}_{ij}(k,P,S)
    (\slashed{k}+m_q)_{jl}^{ }
    = {\rm tr}\biggl[
    \Gamma \phi^{\bar q}(k,P,S) (\slashed{k}+m_q)\biggr]
    = 0\,.
\ea
It is convenient to introduce the notation for the identities
\be\label{Eq:qbar-corr-00}
     {\rm Tr}^{\bar q}[\Gamma] \equiv \frac14\;
     {\rm tr}\,\biggl[\Gamma\,\Phi^{\bar q}(k,P,S)\,(\slashed{k}+m_q)\biggr] = 0\,,
\ee
where $\Gamma$ can be again any Dirac matrix. Proceeding 
analog to the quark correlator case, we obtain from
(\ref{Eq:qbar-corr-00}) the following relations 
among the antiquark amplitudes,
\ba
    A_1^{\bar q} = -\,\frac{m_q}{M}\,A_3^{\bar q}, \quad
    A_2^{\bar q} = 0, \quad
    A_4^{\bar q} = 0, \quad
    A_5^{\bar q} = 0, \quad
    A_6^{\bar q} = -\frac{m_q}{M}\,A_{10}^{\bar q},  
    \label{Eq:ampl-01-bar}\\
    \nonumber\\
    A_7^{\bar q} = \frac{m_q}{M}\,A_{11}^{\bar q}, \quad
    A_9^{\bar q} = 0, \quad 
    A_{10}^{\bar q} = \frac{(P\cdot k)}{M^2}\,A_{11}^{\bar q} +\frac{m_q}{M}\,A_8^{\bar q} \,, \quad
    A_{12}^{\bar q} = 0.
    \label{Eq:ampl-02-bar}
\ea
Analogously to the quark case, also the T-odd antiquark
amplitudes $A_4^{\bar q}$, $A_5^{\bar q}$, $A_{12}^{\bar q}$
vanish and so do the T-even amplitudes $A_2^{\bar q}$ and 
$A_9^{\bar q}$. All non-zero amplitudes are related 
in one or another way to three amplitudes $A_3^{\bar q}$,
$A_8^{\bar q}$, $A_{11}^{\bar q}$ which remain unconstrained
by the equations of motion. The relations of the antiquark
amplitudes in 
(\ref{Eq:ampl-01-bar},~\ref{Eq:ampl-02-bar}) 
resemble those in the quark case in
(\ref{Eq:ampl-02},~\ref{Eq:ampl-04})
except that the current quark mass $m_q$ enters with
opposite sign.

\section{\boldmath Quark correlator in massive case}
\label{Sec-4:q-correlator-massive}

Inserting the relations in (\ref{Eq:ampl-02},~\ref{Eq:ampl-04})
in the quark correlator in (\ref{Eq:correlator-decompose}) we obtain
\be\label{Eq:correlator-decompose-3}
     \Phi^q(k,P,S) = \Phi^q(k,P,S)_{\rm unp} + \Phi^q(k,P,S)_{\rm pol}
\ee
where
\ba\label{Eq:correlator-decompose-4}
    \Phi^q(k,P,S)_{\rm unp} 
    &=& (\slashed{k}+m_q)\,A_{\rm unp}^q \quad \mbox{with} \quad
    A_{\rm unp}^q = A_3^q\,,\nonumber\\
    \Phi^q(k,P,S)_{\rm pol}
    &=& (\slashed{k}+m_q)\gamma_5 \biggl\{
         \slashed{P}\,\frac{(k\cdot S)}{M^2} A_{11}^q 
        -\slashed{S}\,\frac{(P\cdot k)\,A_{11}^q -m_qM\,A_8^q}{M^2} 
       + \frac{(k\cdot S)}{M}\; A_8^q \biggr\}\,.
\ea
In the parton model $\slashed{k}\slashed{k}=k^2=m_q^2$
and the last term in the curly bracket of
(\ref{Eq:correlator-decompose-4}) can be
written for $m_q\neq0$ as 
\be\label{eq:step-towards-w}
     (\slashed{k}+m_q)\gamma_5 \,\frac{(k\cdot S)}{M}\; A_8^q 
   = - (\slashed{k}+m_q)\gamma_5 \,\frac{\slashed{k}}{m_q}\,
     \frac{(k\cdot S)}{M}\; A_8^q \,.
\ee
Inserting the relation (\ref{eq:step-towards-w}) into 
(\ref{Eq:correlator-decompose-4}) allows us to write 
the polarization-dependent part of the correlator as
\ba\label{Eq:correlator-decompose-5}
    \Phi^q(k,P,S)_{\rm pol}
    = (\slashed{k}+m_q)\gamma_5\,\slashed{w}\,A_{\rm pol}^q \;\;
       \ea
where we introduce the polarized amplitude $A_{\rm pol}^q$
and the axial 4-vector $w_q^\mu$ defined as 
\bsub\label{Eq:Apol-w-quark-case}
\ba\label{Eq:pol-vector-1}
     A_{\rm pol}^q &=& -\,\frac{(P\cdot k)\,A_{11}^q -m_qM\,A_8^q}{M^2}\,,\\
     w_q^\mu  &=& 
       S^\mu 
     - P^\mu \frac{(k\cdot S)\,A_{11}^q}{(P\cdot k)\,A_{11}^q -m_qM\,A_8^q} 
     + k^\mu\frac{M}{m_q}\,
     \frac{(k\cdot S)A_8^q}{(P\cdot k)\,A_{11}^q -m_qM\,A_8^q}\;.
\ea\esub
We note that for $w^\mu_q$ to be well-defined in
(\ref{Eq:pol-vector-1})
it must be $m_q\neq 0$ (which is the case in this section) 
and the condition $(P\cdot k)\,A_{11}^q -m_qM\,A_8^q\neq 0$  
must hold. 
We will follow up on this shortly.

\subsection{Quark polarization vector}
\label{Sec:q-correlator-massive-wq}

The 4-vector $w_q^\mu$ has the following important properties.
It has the transformation properties of a polarization vector, 
i.e.\ of an axial 4-vector, and satisfies the condition
\be\label{Eq:pol-vector-2}
      w_q\cdot k = 0\,.
\ee 
These 2 properties are necessary conditions for $w^\mu_q$ 
to be a candidate expression for a quark polarization vector. 
Remarkably, the condition $w_q\cdot k = 0$ holds for any $A_8^q$ 
and $A_{11}^q$ without imposing a relation between these 
amplitudes. 

In order to be a quark polarization vector, 
$w^\mu_q$ must in addition satisfy the requirements
\begin{alignat*}{10}
    \mbox{Condition (A)}\quad & -1 < \; &w_q^2&  <  0 & 
        \mbox{mixed-spin state,}\nonumber\\
    \mbox{Condition (B)}\quad &         &w_q^2&  = -1 \quad &
        \mbox{pure-spin state.}\nonumber
\end{alignat*}
The square of the quark polarization vector is given by 
\be\label{Eq:pol-vec-square}
      w_q^2 = S^2 +
      \frac{(k\cdot S)^2M^2}{[(P\cdot k)\,A_{11}^q -m_qM\,A_8^q]^2}\,
      \biggl((A_{11}^q)^2-(A_{8}^q)^2\biggr)\,.
\ee
In the following, we will explore both possibilities
(A) to which we will refer as the mixed-spin version 
of the model, and (B) to which we will refer as the 
pure-spin version of the model.

\subsection{Mixed-spin model}
\label{Sec:q-correlator-massive-mixed-spin}

Let us first discuss the mixed-spin version of the
model which requires $-1 < \omega_q^2 < 0$.
The condition $\omega^2_q>-1$ is equivalent to
$|A^q_{11}|>|A^q_8|$.
The condition $\omega^2_q<0$ is satisfied as long as 
$(P\cdot k\;A^q_{11}-m_qMA^q_8)^2>0$ which is always the 
case because we had to assume 
in (\ref{Eq:pol-vector-1}) that $(P\cdot k)
A^q_{11}-m_qMA^q_8\neq0$. 
Thus, in the mixed-spin case, besides the inequality
$|A^q_{11}|>|A^q_8|$, the amplitudes $A_8^q$ and 
$A_{11}^q$ remain unrelated.

In this approach, the quark correlator is described in terms of 
3 independent amplitudes, namely $A^q_3$, $A^q_8$, $A^q_{11}$, 
i.e.\ in terms of one unpolarized amplitude and two polarized 
amplitudes: one chiral even and one chiral odd. 

\subsection{Pure-spin state model}
\label{Sec:q-correlator-massive-pure-spin}

If we demand the quarks to be in a pure-spin state, 
$w^\mu_q$ must be normalized as $w^2_q = -1$ in
(\ref{Eq:pol-vec-square}).
This means that the second term on the right-hand-side of 
(\ref{Eq:pol-vec-square}) must vanish which implies the following
condition 
\bsub\label{Eq:q-solutions-options}
\be\label{Eq:q-solution-w2=1}
    w_q^2 
    = -1 \quad \Leftrightarrow \quad
    A_{11}^q = \pm\,A_{8}^q  \,.
\ee
The 2 solutions in (\ref{Eq:q-solution-w2=1}) lead to 2 
different solutions for $A^q_{\rm pol\pm}$ and $w^\mu_{q\pm}$.
One way of writing the pertinent solutions consists in 
eliminating the chiral odd amplitude $A_{11}^q$ which yields
\ba\label{Eq:q-solution-Apol-w}
A_{\rm pol \pm}^q &=& 
    -\,\frac{(\pm P\cdot k)-m_qM}{M^2}\,\,A_8^q
    \,,\label{solutions-Apol-q-pm}\\
w^\mu_{q\pm}  &=&
    S^\mu 
     - \frac{(\pm k\cdot S)}{(\pm P\cdot k)-m_qM}\,P^\mu 
     + \frac{M}{m_q}\,\frac{(\pm k\cdot S)}{(\pm P\cdot k) - m_qM}\,
     (\pm k^\mu)\;.\label{solutions-w-q-pm}
\ea
\esub
This is not the most economic notation, but we have chosen
it to avoid the usage of $\mp$ and in this way the $\pm$ 
always appear together with $k$.
We will show below in Sec.~\ref{Sec-6:TMD-q} that one of 
the 2 solutions is physical, and the other one is unphysical. 
Before that, however, we will discuss the massless case.

\section{\boldmath Quark correlator in the massless case}
\label{Sec-5:q-correlator-massless}

When $m_q=0$, the analysis of the previous section cannot 
be carried out because for massless quarks a polarization 
vector $w_q^\mu$ cannot be defined, see e.g.\ 
\cite{Barone:2001sp,Collins-book}. This can be seen directly 
in (\ref{Eq:q-solution-Apol-w}) where for $m_q\to0$ one 
would encounter a $1/m_q$-singularity. 
In the case $m_q=0$, one has to proceed in a different way 
as shown below.

The relations derived from the equations of motion in
(\ref{Eq:ampl-02},~\ref{Eq:ampl-04}) are of course 
valid also for $m_q=0$. Inserting these relations for 
$m_q=0$ in the quark correlator in  (\ref{Eq:correlator-decompose})
yields an unpolarized and a polarized contribution to 
the quark correlator in  (\ref{Eq:correlator-decompose-3}) 
which are given by
\ba\label{Eq:correlator-decompose-4-m=0}
    \Phi^q(k,P,S)_{\rm unp} 
    &=& \slashed{k}\,A_{\rm unp}^q \,,\nonumber\\
    \Phi^q(k,P,S)_{\rm pol}
    &=& \slashed{k}\,\gamma_5 \biggl\{
        -\lambda\; 
        +\slashed{b}_T
       \biggr\}\,A^q_{\rm pol} \,
\ea
where, assuming $A^q_8\neq0$, we defined 
\be\label{Eq:pol-def-massless-case}
    A_{\rm unp}^q = A_3^q, \quad
    A^q_{\rm pol} = -\,\frac{(P\cdot k)}{M^2}\;A_8^q\,, \quad 
    \lambda =  \frac{M(k\cdot S)}{(P\cdot k)} \, , \quad
    b_T^\mu  = \biggl(S^\mu-P^\mu\,\frac{(k\cdot S)}{(P\cdot k)}\biggr)
    \,\frac{A_{11}^q }{A_8^q}\,.
\ee
The quantities $\lambda$ and $b_T^\mu$ are defined in
(\ref{Eq:pol-def-massless-case}) following the common 
conventions in literature, see for instance (2.3.8) 
in Ref.~\cite{Barone:2001sp} or Appendix A of
Ref.~\cite{Collins-book}.
Their properties are discussed in the next section.

\subsection{Spin density matrix of massless quarks}

The quantities $\lambda$ and $b_T^\mu$ in
(\ref{Eq:pol-def-massless-case}) 
have the following properties.

\begin{enumerate} [label=(\roman*)]
    \item In the high-energy limit when the nucleon moves very fast, 
i.e.\ $P^\mu\to\infty$, and is polarized along its direction of
motion, the nucleon polarization vector is given by 
$S^\mu = \lambda_N P^\mu/M + \dots$
where $\lambda_N=\pm1$ is (twice) the helicity of the nucleon 
\cite{Barone:2001sp} 
(see Sec.~\ref{Subsec:discussion-other-approaches} for 
a reminder) and the dots indicate
terms suppressed in the high-energy limit. In such an 
infinite momentum frame, the quantity $\lambda$ defined 
in (\ref{Eq:pol-def-massless-case}) is
given by $\lambda = \lambda_N$, i.e.\ depending on the sign 
of $A^q_{\rm pol}$, the massless quark has the same or opposite
helicity as the nucleon. Thus in the infinite momentum frame,
$\lambda$ has the expected intuitive interpretation as
(twice) the quark helicity.

\item In the general case, a quark may of course have transverse
polarization. In the massless case, this is described by the
vector $b_T^\mu$ which is transverse with respect to the quark
momentum $k^\mu$. 
It is important to stress that $b_T^\mu$ introduced in
(\ref{Eq:pol-def-massless-case}) has the property 
$k\cdot b_T = 0$ for any $A_8^q$ and $A_{11}^q$.

\item  Also in the massless case, we can distinguish pure-spin 
and  mixed-spin states. They are defined as follows:
\begin{alignat}{10}
\text{Condition}\hspace{.15cm} \mbox{(A)} \quad & -1 < \; & b_T^2-\lambda^2&  <  0 & \mbox{mixed-spin state,}
\nonumber\\
\text{Condition}\hspace{.15cm}\mbox{(B)} \quad && b_T^2-\lambda^2&  = -1 \quad & \mbox{pure-spin state.} \label{Eq:cond-A-B}
\end{alignat}

\end{enumerate}

\subsection{Massless mixed-spin model}

If one chooses to work with massless quarks in a mixed-spin
state, then
$-1 < b_T^2- \lambda^2 <  0$ must hold. 
The condition $b_T^2-\lambda^2>-1$ will always be satisfied as long as
$|A_{11}^q|>|A_8^q|$. The condition $b_T^2- \lambda^2 <  0$ is trivial
and always satisfied without imposing any new requirements. 
Recalling that we had to exclude the case $A_8^q=0$ from the
very beginning, we conclude that the mixed-spin parton model
is consistently defined provided $0<|A_8^q|<|A_{11}^q|$. 
Similarly to the massive case, in this approach the quark correlator 
is described in terms of 3 independent amplitudes, namely
the unpolarized $A^q_3$, chiral even polarized $A^q_8$, and 
chiral odd polarized $A^q_{11}$ amplitude.
In practice, these 3 independent amplitudes can be determined from, 
for instance $f_1^q(x)$, $g_1^q(x)$, $h_1^q(x)$ at some scale which 
is part of the model. This corresponds to the parton model 
version discussed in Ref.~\cite{DAlesio:2009cps}.

\subsection{Massless pure-spin model}

If we choose to work with a parton model of massless quarks
in a pure-spin state, then the condition (B) in (\ref{Eq:cond-A-B}) implies
\be\label{Eq:sol-pol-massless}
     b_T^2-\lambda^2  
     = 
     -\,\frac{M^2(k\cdot S)^2}{(P\cdot k)^2}\biggl(
     1- \frac{(A_{11}^q)^2}{(A_8^q)^2}\biggr)+ S^2 
     = -1\,
     \quad \Leftrightarrow \quad
     A_8^q = \pm\,A_{11}^q\,.
\ee
Also in the massless case, we have two choices  for a
pure-spin state parton model, one of which will turn out to
be physical and the other unphysical. We will
discuss this in the next section.

\section{Quark TMD\lowercase{s}}
\label{Sec-6:TMD-q}

In this section we will discuss the results for TMDs 
following from the quark correlator in the parton model
derived in the previous sections. We will keep $m_q\neq 0$
and comment on the massless case where necessary.

\subsection{Unpolarized TMDs}

The results for the T-even unpolarized TMDs are of course
independent of the (mixed-spin or pure-spin state)
polarization of partons, and can be uniquely expressed 
in terms of the amplitude $A_3^q$ in the following way
\bsub\label{Eq:rel-unp-q}
\ba
    f_1^q(x,k_T) &=& 
        2P^+\int dk^- \biggl[xA_3^q(P\cdot k)\biggr]_{k^+=xP^+},
        \label{Eq:f1-q-final}\\
    f^{\perp q}(x,k_T) &=& 
        2P^+\int dk^- \biggl[A_3^q(P\cdot k)\biggr]_{k^+=xP^+},\\
    e^q(x,k_T) &=& 
        2P^+\int dk^- \biggl[\frac{m_q}{M}\,
        A_3^q(P\cdot k)\biggr]_{k^+=xP^+}.
\ea\esub
Here and in the following we shall abbreviate 
the notation for the amplitudes as 
$A_3^q(P\cdot k,\,k^2)=A_3^q(P\cdot k)$ because
in the parton model $k^2=m_q^2$ is fixed 
(but we will reinstate the notation 
$A_3^q(P\cdot k,\,k^2)$ when it will become important
to stress the specific $k^2$ dependence in the parton
model).

We see that $e^q(x,k_T)$ is proportional to the current 
quark mass and becomes zero if one considers the parton 
model with massless quarks. This is the only TMD in the 
parton model with this property.

\subsection{Polarized TMDs for partons in mixed-spin state}

The T-even chiral even polarized TMDs in the
mixed-spin state parton model with massive quarks 
are expressed in terms of the chiral even amplitude
$A_8^q$ and the chiral odd amplitude $A_{11}^q$ 
entering as current quark mass effect. 
The model expressions are given by 
\bsub\label{Eq:mixed-spin-TMDq-all}
\ba
    g_1^q(x,k_T) &=& 2P^+\int dk^-\Biggl[
    \frac{x^2M^2 -x \,P\cdot k+m_q^2}{M^2} \,A_8^q(P\cdot k)
    -\frac{m_q}{M}\,x\,
    A_{11}^q(P\cdot k)\Biggr]_{k^+=xP^+}, 
    \label{Eq:mixed-spin-TMDq-g1}\\
    g_{1T}^{\perp q}(x,k_T) &=& 2P^+\int dk^- 
    \Biggl[xA_8^q(P\cdot k)-\dfrac{m_q}{M}
    A_{11}^q(P\cdot k)\Biggr]_{k^+=xP^+},\label{OLD-Eq-ms-g1Tperpq}\\
    g_T^q(x,k_T) &=& 2P^+\int dk^- \Biggl[
    \dfrac{\vec{k}_T^2+2m_q^2}{2M^2}\,A_8^q(P\cdot k)
    -\dfrac{m_q}{M}\;\dfrac{P\cdot k}{M^2}\,A_{11}^q(P\cdot k)
    \Biggr]_{k^+=xP^+},\label{OLD-Eq-ms-gTq}\\
    g_L^{\perp q}(x,k_T) &=& 2P^+\int dk^- \Biggl[
    \frac{x\,M^2-P\cdot k}{M^2}\,A_8^q(P\cdot k)\Biggr]_{k^+=xP^+},\label{OLD-Eq-ms-gLperpq}\\
    g_T^{\perp q}(x,k_T) &=& 2P^+\int dk^- \Biggl[A_8^q(P\cdot k)\Biggr]_{k^+=xP^+}.\label{OLD-Eq-ms-gTperpq}
\ea\esub
For the T-even chiral odd polarized TMDs the situation
is opposite: they are given in terms of the 
chiral odd amplitude $A_{11}^q$ while the chiral even
amplitude $A_8^q$ enters as current
quark mass effect. The model expressions read 
\bsub\begin{eqnarray}
    h_1^q(x,k_T) &=& 2P^+\int dk^- \Biggl[
    \dfrac{\vec{k}_T^2-2\,x\,P\cdot k}{2M^2}\,A_{11}^q(P\cdot k)
    +x\,\dfrac{m_q}{M}\,A_8^q(P\cdot k) \Biggr]_{k^+=xP^+},\\
    h_{1L}^{\perp q}(x,k_T) &=& 2P^+\int dk^- \Biggl[
    xA_{11}^q(P\cdot k) -\dfrac{m_q}{M}A_8^q(P\cdot k)\Biggr]_{k^+=xP^+},\\
    h_{1T}^{\perp q}(x,k_T) &=& 2P^+\int dk^- \Biggl[
    A_{11}^q(P\cdot k)\Biggr]_{k^+=xP^+},\\
    h_L^q(x,k_T) &=& 2P^+\int dk^- \Biggl[
    \frac{x^2M^2-2\,x\,P\cdot k}{M^2}\,A_{11}^q(P\cdot k)
    +\dfrac{m_q}{M}\;\dfrac{P\cdot k}{M^2}\,A_8^q(P\cdot k)
    \Biggr]_{k^+=xP^+},\\
    h_T^q(x,k_T) &=& 2P^+\int dk^- \Biggl[
    \frac{x\,M^2-P\cdot k}{M^2}\,A_{11}^q(P\cdot k)\Biggr]_{k^+=xP^+},\\
    h_T^{\perp q}(x,k_T) &=& 
    2P^+\int dk^- \Biggl[-\dfrac{P\cdot k}{M^2} A_{11}^q(P\cdot k)+\dfrac{m_q}{M}A_8^q(P\cdot k)\Biggr]_{k^+=xP^+}.
\end{eqnarray}\esub
For massless partons the situation simplifies. Then all 
chiral even polarized TMDs are expressed in terms of the 
amplitude $A_8^q$, while all chiral odd polarized TMDs are expressed in terms of the amplitude
$A_{11}^q$.

\subsection{Polarized TMDs for partons in pure-spin state}

In the pure-spin state parton model all polarized 
(chiral even {\it and} chiral odd) TMDs, can be expressed in terms 
of one single amplitude. We choose $A_8^q$ for that, and replace 
$A_{11}^q$ by $A_{11}^q=\pm A_8^q$. In the following the upper
(lower) sign in $(\pm)$ is associated with the upper
(lower) sign in the two solutions $w^q_{\pm}$ for the quark 
polarization vector and polarized amplitude $A^q_{\rm pol\pm}$
which correspond to the choices $A_{11}^q=\pm A_8^q$.
For massive partons in a pure-spin state, the model expressions 
for the chiral even polarized TMDs are given by

\bsub\ba
    g_1^q(x,k_T) &=& 
    2P^+\int dk^- \Biggl[
    \frac{x^2M^2-x\,P\cdot k+m_q^2}{M^2}\,A_8^q(P\cdot k)
    -\dfrac{m_q}{M}\,x\,\biggl[\pm A_8^q(P\cdot k)\biggr]
    \Biggr]_{k^+=xP^+},
    \\
    g_{1T}^{\perp q}(x,k_T) &=& 2P^+\int dk^-\Biggl[
    x\,A_8^q(P\cdot k) -  
    \dfrac{m_q}{M}\biggl[\pm A_8^q(P\cdot k)\biggr]
    \Biggr]_{k^+=xP^+},\\
    g_T^q(x,k_T) &=& 2P^+\int dk^-\Biggl[
    \dfrac{\vec{k}_T^2+2m_q^2}{2M^2}\,A_8^q(P\cdot k)
    -\dfrac{m_q}{M}
    \biggl[\pm\,\dfrac{P\cdot k}{M^2}\, A_8^q(P\cdot k)\biggr]\Biggr]_{k^+=xP^+},
\ea\esub
where we do not quote the expressions for $g_L^{\perp q}(x,k_T)$ 
and $g_T^{\perp q}(x,k_T)$ since they do not depend on $A_{11}^q$ and 
are the same as in (\ref{OLD-Eq-ms-gLperpq},~\ref{OLD-Eq-ms-gTperpq}).
Since $m_q\ll M$, the difference between the two solutions for 
$w^q_{\pm}$ and $A^q_{\rm pol\pm}$ is in practice numerically 
negligibly small, and disappears in the massless case. In fact, 
in the massless case the mixed-spin and pure-spin versions of 
the parton model give exactly the same results for the chiral even
polarized TMDs. 

The model expressions for the polarized chiral odd 
TMDs are given by 
\bsub\label{Eqs:TMDq-pure-spin-plus-minus}\begin{eqnarray}
    h_1^q(x,k_T) &=& 2P^+\int dk^- \Biggl[
    \pm\,\dfrac{\vec{k}_T^2-2\,x\,P\cdot k}{2M^2}
    \,A_8^q(P\cdot k)
    +\dfrac{m_q}{M}\,x\,A_8^q(P\cdot k)
    \Biggr]_{k^+=xP^+},\\
    h_{1L}^{\perp q}(x,k_T) &=& 2P^+\int dk^- \Biggl[
    \pm\,x\,A_8^q(P\cdot k)
    - \dfrac{m_q}{M}\,A_8^q(P\cdot k)\Biggr]_{k^+=xP^+},\\
    h_{1T}^{\perp q}(x,k_T) &=& 2P^+\int dk^- \Biggl[
    \pm\,A_{8}^q(P\cdot k)
    \Biggr]_{k^+=xP^+},\\
    h_L^q(x,k_T) &=& 2P^+\int dk^- \Biggl[
    \pm\,\frac{x^2M^2-2\,x\,P\cdot k}{M^2}\,
    A_{8}^q(P\cdot k)
    +\dfrac{m_q}{M}\,
    \dfrac{P\cdot k}{M^2}\,A_{8}^q(P\cdot k)
    \Biggr]_{k^+=xP^+},\\
    h_T^q(x,k_T) &=& 2P^+\int dk^- \Biggl[
    \pm \, \frac{x\,M^2-P\cdot k}{M^2}\,
    A_{8}^q(P\cdot k)
    \Biggr]_{k^+=xP^+},\\
    h_T^{\perp q}(x,k_T) &=& 
    2P^+\int dk^- \Biggl[
    \pm\,\biggl(-\,
    \dfrac{P\cdot k}{M^2}\,A_{8}^q(P\cdot k)\biggr)
    +\dfrac{m_q}{M}\, A_{8}^q(P\cdot k)
    \Biggr]_{k^+=xP^+}.
\end{eqnarray}\esub
If we consider that quark mass effects are negligible 
in practical applications, we notice that the signs $\pm$
associated with the solutions for $w^q_{\pm}$ and 
$A_{\rm pol\pm}^q$ yield different predictions for the 
overall signs of the chiral odd polarized TMDs. 
For massless quarks, the two choices give exactly 
opposite signs for chiral odd polarized TMDs.

The negative-sign solution 
for $w^q_{\pm}$ and $A_{\rm pol\pm}^q$ in 
(\ref{Eq:q-solutions-options}) 
coincides with the conventions in 
\cite{Zavada:2001bq,Zavada:2002uz,Efremov:2004tz,
Zavada:2007ww,Efremov:2009ze,Zavada:2009ska,Efremov:2009vb,
Efremov:2010mt,Zavada:2011cv,Zavada:2013ola,Zavada:2015gaa,
Zavada:2019yom,Bastami:2020rxn} 
and predicts transversity and helicity quark TMDs to 
have equal signs in agreement with other 
models and lattice QCD 
\cite{
Jaffe:1991ra,Ioffe:1994aa,Pobylitsa:1996rs,
Signal:1996ct,Jakob:1997wg,Scopetta:1997qg,
Suzuki:1997vu,Ma:1997gy,Suzuki:1997wv,
Wakamatsu:1998rx,Gamberg:1998vg,Schweitzer:2001sr,
Mukherjee:2001zx,Meissner:2007rx,Cloet:2007em,Avakian:2008dz,
Pasquini:2008ax,Bacchetta:2008af,She:2009jq,Avakian:2009jt,Avakian:2010br,
Chen:2016utp,Gutsche:2016gcd,Kofler:2017uzq,Maji:2017bcz,
Signal:2021aum,Alexandrou:2021oih,Alexandrou:2018eet,Alexandrou:2021bbo,HadStruc:2021qdf}.\footnote{We 
    remark that one can determine the physical sign 
    solution for $h_1^q(x)$ also from model and 
    lattice QCD studies of the nucleon tensor charge 
    $\int_0^1dx(h_1^q-h_1^{\bar q})(x)$ 
    \cite{Kim:1995bq,Aoki:1996pi,Gockeler:2005cj,QCDSF:2006tkx,Alexandrou:2019ali,Bali:2018zgl,Alexandrou:2022dtc,Aoki:2021kgd,Lorce:2007fa,Ledwig:2010tu},
    under the safe assumption that the sign of the tensor 
    charge is determined by $h_1^q(x)$.}
We therefore conclude that the negative-sign solution in 
(\ref{Eq:q-solutions-options}) is physical. 
From the positive-sign solution $w^q_+$ and 
$A_{\rm pol+}^q$ in 
(\ref{Eq:q-solutions-options}),
one would obtain transversity and helicity quark TMDs
with opposite signs. This has not been observed in 
any model or lattice study we are aware of. 
We therefore discard the positive-sign solution in 
(\ref{Eq:q-solutions-options}) as unphysical.

It is natural to encounter two solutions because 
the pure-spin state condition $w_q^2=-1$ is a 
quadratic equation. Interestingly, the pure-spin
state parton model by itself cannot distinguish which 
solution is physical and which one is unphysical, and 
thus does not predict the sign of transversity and other
chiral odd TMDs. We have to determine the physical 
solution using results from other models or lattice QCD as 
a guideline.\footnote{%
    Let us recall that the sign of $h_1^a(x)$ cannot 
    be inferred from experiment because chiral odd 
    (TMDs, fragmentation, etc) functions  
    enter observables paired with other
    chiral odd functions. The positive sign of 
    $h_1^u(x)$ in parametrizations (see, e.g., 
    \cite{Cammarota:2020qcw} for a recent extraction)
    is a convention based models and lattice QCD
    predictions
    \cite{
    Jaffe:1991ra,Ioffe:1994aa,Pobylitsa:1996rs,
    Signal:1996ct,Jakob:1997wg,Scopetta:1997qg,
    Suzuki:1997vu,Ma:1997gy,Suzuki:1997wv,
    Wakamatsu:1998rx,Gamberg:1998vg,Schweitzer:2001sr,
    Mukherjee:2001zx,Meissner:2007rx,Cloet:2007em,Avakian:2008dz,
    Pasquini:2008ax,Bacchetta:2008af,She:2009jq,Avakian:2009jt,Avakian:2010br,
    Chen:2016utp,Gutsche:2016gcd,Kofler:2017uzq,Maji:2017bcz,
    Signal:2021aum,Alexandrou:2021oih,Alexandrou:2018eet,Alexandrou:2021bbo,HadStruc:2021qdf,
    Kim:1995bq,Aoki:1996pi,Gockeler:2005cj,
    QCDSF:2006tkx,Alexandrou:2019ali,Bali:2018zgl,Alexandrou:2022dtc,Aoki:2021kgd,Lorce:2007fa,Ledwig:2010tu}.}

Thus, the physical (negative-sign) solution corresponds 
to the covariant parton model in
\cite{Zavada:2001bq,Zavada:2002uz,Efremov:2004tz,
Zavada:2007ww,Efremov:2009ze,Zavada:2009ska,Efremov:2009vb,
Efremov:2010mt,Zavada:2011cv,Zavada:2013ola,Zavada:2015gaa,
Zavada:2019yom,Bastami:2020rxn} and is given by
\ba\label{Eq:q-solution-Apol-w-phys}
    A_8^q           &=& - A_{11}^q \, , \quad
    A_{\rm pol -}^q \;=\; \frac{(P\cdot k)+ m_qM}{M^2}\;
                        A_8^q\,,\nonumber\\
    w^\mu_{q-}      &=& 
    S^\mu 
     - \frac{(k\cdot S)}{(P\cdot k)+m_qM}\;P^\mu
     - \frac{M}{m_q}\,\frac{(k\cdot S)}{(P\cdot k)+m_qM}\;k^\mu\;.
\ea
The final model expressions for the polarized TMDs 
in the pure-spin state version of the model are given by
\bsub\label{Eq:pure-spin-TMDq-all}
\ba
    g_1^q(x,k_T) &=& 
    2P^+\int dk^- \Biggl[
    \frac{x^2M^2-x\,P\cdot k+x\,m_qM+m_q^2}{M^2}\,A_8^q(P\cdot k)
    \Biggr]_{k^+=xP^+}, \\
    g_{1T}^{\perp q}(x,k_T) &=& 2P^+\int dk^-\Biggl[
    \frac{x\,M+m_q}{M}\;A_8^q(P\cdot k)\Biggr]_{k^+=xP^+},\\
    g_T^q(x,k_T) &=& 2P^+\int dk^-\Biggl[
    \dfrac{M\,\vec{k}_T^2+2\,m_q\,P\cdot k+2m_q^2M}{2M^3}\,A_8^q(P\cdot k)
    \Biggr]_{k^+=xP^+}, \\
    g_L^{\perp q}(x,k_T) &=& 2P^+\int dk^- \Biggl[
    \frac{x\,M^2-P\cdot k}{M^2}\,A_8^q(P\cdot k)\Biggr]_{k^+=xP^+},
    \label{OLD2-Eq-ms-gLperpq}\\
    g_T^{\perp q}(x,k_T) &=& 2P^+\int dk^- \Biggl[A_8^q(P\cdot k)
    \Biggr]_{k^+=xP^+}. \\
    h_1^q(x,k_T) &=& 2P^+\int dk^- \Biggl[
    \dfrac{2\,x\,P\cdot k-\vec{k}_T^2+2xm_qM}{2M^2}\,A_8^q(P\cdot k)
    \Biggr]_{k^+=xP^+},\\
    h_{1L}^{\perp q}(x,k_T) &=& 2P^+\int dk^- \Biggl[\, -
    \frac{x\,M+m_q}{M}\;A_8^q(P\cdot k)\Biggr]_{k^+=xP^+},\\
    h_{1T}^{\perp q}(x,k_T) &=& 2P^+\int dk^- \Biggl[\,-\,
    A_{8}^q(P\cdot k)\Biggr]_{k^+=xP^+},\\
    h_L^q(x,k_T) &=& 2P^+\int dk^- \Biggl[
    \frac{2\,x\,M\,P\cdot k-x^2M^3+m_q\,P\cdot k}{M^3}
    \,A_{8}^q(P\cdot k)\Biggr]_{k^+=xP^+},\\
    h_T^q(x,k_T) &=& 2P^+\int dk^- \Biggl[
    \frac{P\cdot k-x\,M^2}{M^2}\,A_{8}^q(P\cdot k)\Bigr]
    \Biggr]_{k^+=xP^+},\\
    h_T^{\perp q}(x,k_T) &=& 
    2P^+\int dk^- \Biggl[
    \dfrac{P\cdot k+m_qM}{M^2}\,A_{8}^q(P\cdot k)\Biggr]_{k^+=xP^+}.
\end{eqnarray}
\esub

Above, we explored theoretical studies of
$h_1^q$ to determine the physical sign solution, 
because transversity is arguably the most widely studied 
chiral odd nucleon distribution function. 
For completeness, let us remark that in principle we could
have used any other chiral odd quark distribution function,
for instance $h_L^q(x)$ for which some model and lattice 
studies are available in literature \cite{
    Jaffe:1991ra,Signal:1996ct,Jakob:1997wg,
    Avakian:2010br,Dressler:1999hc,Wakamatsu:2000fd,
    Bhattacharya:2021moj}.
One obtains the same result from $h_L^q(x)$.
It is important to remark that there is more than 
one way to determine the physical sign --- including
further possibilities to use chiral odd TMDs like 
pretzelosity $h_{1T}^{\perp q}$, the gear-worm function
$h_{1L}^{\perp q}$, or other polarized chiral odd TMDs.
The fact that one concludes unanimously the same 
relative (negative) sign between the amplitudes
$A_8^q$ and $A_{11}^q$ using the different methods supports 
the consistency of the approach.

\section{Antiquarks in the parton model}
\label{Sec-7:correlator-barq}

The discussion of the antiquark correlator parallels
that of the quark correlator such that we can abbreviate
many details and focus on showing the main results.

\subsection{Correlator for massive antiquarks}

Inserting the relations in (\ref{Eq:ampl-02-bar}) in 
the antiquark correlator in (\ref{Eq:correlator-qbar}) 
and performing exactly the same steps as in the polarized 
correlator as in (\ref{eq:step-towards-w}) we obtain
\bsub\be\label{Eq:correlator-decompose-3-bar}
     \Phi^{\bar q}(k,P,S) = \Phi^{\bar q}(k,P,S)_{\rm unp} + \Phi^{\bar q}(k,P,S)_{\rm pol}
\ee
with
\ba\label{Eq:correlator-decompose-4-bar}
    \Phi^{\bar q}(k,P,S)_{\rm unp} 
    &=& (\slashed{k}-m_q)\,A_{\rm unp}^{\bar q} 
    \; ,  \\
\label{Eq:correlator-decompose-5-bar}
    \Phi^{\bar q}(k,P,S)_{\rm pol}
    &=&
    (\slashed{k}-m_q)\gamma_5\,\slashed{w}_{\bar q}
    \,A_{\rm pol}^{\bar q}  \;\;
    \ea
where we introduced $A_{\rm unp}^{\bar q}$,
$A_{\rm pol}^{\bar q}$, $w^\mu_{\bar q}$ 
defined as\footnote{We recall that by convention
    antifermions are polarized opposite to the
    polarization vector, cf.\
    Sec.~\ref{Subsec:discussion-other-approaches}.
    We stress that this has no practical relevance 
    for our results and the TMDs of both quarks 
    and antiquarks have their usual partonic 
    interpretations. 
    \label{footnote-w-qbar}}
\ba\label{Eq:new-pol-vector-1-qbar}
    A_{\rm unp}^{\bar q} &=& A_3^{\bar q} \,,\nonumber\\
    A_{\rm pol}^{\bar q} &=& -\,
    \frac{(P\cdot k)\,A_{11}^{\bar q}+m_qM\,A_8^{\bar q} }{M^2}\,,\\
     w^\mu_{\bar q}  &=& 
       S^\mu 
     - P^\mu \frac{(k\cdot S)\,A_{11}^{\bar q} }{(P\cdot k)\,A_{11}^{\bar q}  +m_qM\,A_8^{\bar q} } 
     - k^\mu\frac{M}{m_q}\,
     \frac{(k\cdot S)A_8^{\bar q} }{(P\cdot k)\,A_{11}^{\bar q}  +m_qM\,A_8^{\bar q} }\;.
     \label{Eq:new-pol-vector-1-bar}
\ea\esub
In (\ref{Eq:new-pol-vector-1-bar}) 
we assumed that $m_q\neq0$ and 
$(P\cdot k)\,A_{11}^{\bar q} +m_qM\,A_8^{\bar q} \neq 0$. 
The axial 4-vector $w_{\bar q}$ plays the role of the
antiquark polarization vector. It satisfies 
$k\cdot w_{\bar q}=0$ for any
$A_8^{\bar q}$ and $A_{11}^{\bar q}$. 

In the mixed-spin version of the parton model, 
we have 
\be
    -1 < w^2_{\bar q}< 0
\ee
where the upper bound is always satisfied. The lower bound 
is satisfied provided $|A_{11}^{\bar q}|<|A_8^{\bar q}|$.
In this version of the model, the antiquark correlator 
is described in terms of three independent amplitudes 
$A_3^{\bar q}$, $A_8^{\bar q}$, $A_{11}^{\bar q}$. 

In the pure-spin state version of the model, the 
polarized amplitudes $A_8^{\bar q}$ and $A_{11}^{\bar q}$
are related as
\bsub\label{Eq:qbar-solutions-options}
\be\label{Eq:qbar-solution-w2=1}
    w^2_{\bar q} = -1 \quad \Leftrightarrow \quad
    A_{11}^{\bar q} = \pm A_8^{\bar q}\;.
\ee
The two solutions $A_8^{\bar q}=\pm A_{11}^{\bar q}$ imply
\ba\label{Eq:newer-pol+/-}
     A_{\rm pol\pm}^{\bar q}  &=& -\,\frac{(\pm P\cdot k) + m_qM}{M^2}\,A_8^{\bar q},
     \label{Eq:qbar-solution-Apol-w}\\
     w^\mu_{{\bar q}\pm}  &=& 
       S^\mu 
     - P^\mu \frac{(\pm k\cdot S)}{(\pm P\cdot k) + m_qM} 
     - \frac{M}{m_q}\,
     \frac{(\pm k\cdot S)}{(\pm P\cdot k) + m_qM}\,
     (\pm k^\mu)\;.
     \label{solutions-w-qbar-pm}
\ea\esub

\subsection{\boldmath Correlator for massless antiquarks}
\label{Sec:qbar-correlator-massless}

In the massless case, the unpolarized and 
polarized parts of the antiquark correlator 
in (\ref{Eq:correlator-decompose-4-bar}) 
are described as\footnote{In our notation,
    the helicity of the antiquark is given 
    by $(-\lambda)$, cf.\ footnote 
    \ref{footnote-w-qbar}, the reminder in 
    Sec.~\ref{Subsec:discussion-other-approaches},
    and the ``mental health warning'' in 
    Sec.~6.4 of Ref.~\cite{Collins-book}.
    \label{footnote-lambda-qbar}}
\bsub\ba\label{Eq:correlator-decompose-4-bar-massless}
    \Phi^{\bar q}(k,P,S)_{\rm unp} 
     = \slashed{k}\,A_{\rm unp}^{\bar q} \,,\quad
    \Phi^{\bar q}(k,P,S)_{\rm pol}
     = \slashed{k}\,\gamma_5 \biggl\{
        -\lambda\; 
        +\slashed{b}_T
       \biggr\}\,A^{\bar q}_{\rm pol} \,
\ea
where, assuming $A^{\bar q}_8\neq0$, we introduced 
\be\label{Eq:pol-def-massless-case-qbar}
    A_{\rm unp}^{\bar q} = A_3^{\bar q}, \quad
    A^{\bar q}_{\rm pol} = -\,\frac{(P\cdot k)}{M^2}\;A_8^{\bar q}\,, \quad 
    \lambda =  \frac{M(k\cdot S)}{(P\cdot k)} \, , \quad
    b_T^\mu  = \biggl(S^\mu-P^\mu\,\frac{(k\cdot S)}{(P\cdot k)}\biggr)
    \,\frac{A_{11}^{\bar q} }{A_8^{\bar q}}\,.
\ee\esub
The condition $-1< b_T^2-\lambda^2<0$ defining the massless 
spin state model holds provided $|A_{11}^{\bar q}|<|A_8^{\bar q}|$.
In this case case, the model is characterized in terms of three
independent amplitudes $A_3^{\bar q}$, $A_8^{\bar q}$, 
$A_{11}^{\bar q}$. 

In the pure-spin case for antiquarks, it is 
required that $b_T^2-\lambda^2=-1$ which implies 
$A_{11}^{\bar q}= \pm\,A_8^{\bar q}$. The description 
of the antiquark correlator proceeds then in terms of two 
independent amplitudes which can be chosen to be
$A_3^{\bar q}$, $A_8^{\bar q}$.

\subsection{Unpolarized antiquark TMDs }

The T-even unpolarized antiquark TMDs are
uniquely expressed in terms of the amplitude
$A_3^{\bar q}$ as follows, 
c.f.\ (\ref{Eq:TMDs-qbar-amp}), 
\bsub\label{Eq:rel-unp-qbar}
\ba
    \label{Eq:TMDqbar-f1}
        f_1^{\bar q}(x,k_T) &=& 
        2P^+\int dk^- \biggl[x
        A_3^{\bar q}(P\cdot k)\biggr]_{k^+=xP^+},\\
    \label{Eq:TMDqbar-fperp}
        f^{\perp {\bar q}}(x,k_T) &=& 
        2P^+\int dk^- \biggl[
        A_3^{\bar q}(P\cdot k)\biggr]_{k^+=xP^+},
        \\
    \label{Eq:TMDqbar-e}
        e^{\bar q}(x,k_T) &=& 
        2P^+\int dk^- \biggl[\frac{m_q}{M}\,
        A_3^{\bar q}(P\cdot k)\biggr]_{k^+=xP^+}.
        \label{Eq:rel-eqbar-f1bar}
\ea\esub

\subsection{Polarized chiral even antiquark TMDs in 
mixed-spin case}

The polarized chiral even TMDs in the mixed-spin model
are given by
\bsub
\label{Eq:TMDs-qbar-chiral-even-mixed}
\begin{eqnarray}
    \label{Eq:mixed-spin-TMDqbar-g1}
        g_1^{\bar q}(x,k_T) 
        \; &=& \;
        2P^+ \int dk^- 
        \Biggl[
        \frac{x\,P \cdot k-x^2M^2-m_q^2}{M^2}\,A_8^{\bar q}
        -\frac{m_q}{M}\,x\,A_{11}^{\bar q}
        \Biggr]_{k^+=xP^+}\,, 
        \\
    \label{Eq:mixed-spin-TMDqbar-g1T}
         g_{1T}^{\perp \bar q}(x,k_T) \; &=& \; 
         2P^+ \int dk^-\Biggl[-x\,A_8^{\bar q}
         -\frac{m_q}{M}\,A_{11}^{\bar q}\Biggr]_{k^+=xP^+}\,,
          \\
    \label{Eq:mixed-spin-TMDqbar-gT}
         g_T^{\bar q}(x,k_T) \; &=& \; 
         2P^+ \int dk^- 
         \Biggl[
         -\,\frac{\vec{k}_T^{\,2}+2m_q^2}{2M^2} \; A_8^{\bar q} 
         - \frac{m_q}{M}\;\frac{P\cdot k}{M^2}\,A_{11}^{\bar q} 
         \Biggr]_{k^+=xP^+}\ \, , \\
    \label{Eq:mixed-spin-TMDqbar-gLperp}
         g^{\perp \bar q}_L(x,k_T) \; &=& \;  
         2P^+ \int dk^- 
         \Biggl[-\,\frac{x M^2-P \cdot k}{M^2}\; 
         A_8^{\bar q}\Biggr]_{k^+=xP^+}\ \, , 
        \\
    \label{Eq:mixed-spin-TMDqbar-gTperp}
         g^{\perp \bar q}_T(x,k_T) \; &=& \; 2P^+ \int dk^-
         \Biggl[-\,A_8^{\bar q}\;\Biggr]_{k^+=xP^+}\ \, .
\end{eqnarray}
\esub

\subsection{Polarized chiral odd antiquark TMDs in mixed-spin state case}

The polarized chiral odd TMDs in the mixed-spin model
are given by
\bsub
\label{Eq:TMDs-qbar-chiral-odd-mixed}
\begin{eqnarray}
    \label{Eq:mixed-spin-TMDqbar-h1}
         h_1^{\bar q}(x,k_T) \; &=& \; 
        2P^+ \int dk^- \Biggl[
         \frac{\vec{k}_T^{\,2}-2\,x\,P\cdot k}{2M^2} \; A_{11}^{\bar q} - \frac{m_q}{M}\,x\,A_8^{\bar q}
        \Biggr]_{k^+=xP^+} \, ,\\
    \label{Eq:mixed-spin-TMDqbar-h1Lperp}
         h^{\perp \bar q}_{1L}(x,k_T) \; &=& \; 
         2P^+ \int dk^- \Biggl[
         x\, A_{11}^{\bar q}
         +\frac{m_q}{M}\,A_8^{\bar q}
         \Biggr]_{k^+=xP^+} \, , 
         \\
    \label{Eq:mixed-spin-TMDqbar-h1Tperp}
         h^{\perp \bar q}_{1T}(x,k_T) \; &=& \; 
         2P^+ \int dk^- \Biggr[\;A_{11}^{\bar q}\;
         \Biggr]_{k^+=xP^+} \, , \\
    \label{Eq:mixed-spin-TMDqbar-hL}
        h_L^{\bar q}(x,k_T) \; &=& \; 
        2P^+ \int dk^- 
        \Biggl[
        \frac{x^2M^2-2\,x\,P \cdot k}{M^2} A_{11}^{\bar q}
        - \frac{m_q}{M}\,\frac{P \cdot k}{M^2}\;A_8^{\bar q}
         \Biggr]_{k^+=xP^+} \, , \\
    \label{Eq:mixed-spin-TMDqbar-hTperp}
         h^{\perp \bar q}_T(x,k_T) \; &=& \; 
         2P^+ \int dk^- \Biggl[ - \,
         \frac{P\cdot k}{M^2}\,A_{11}^{\bar q}
         -\frac{m_q}{M}\,A_8^{\bar q}\Biggr]_{k^+=xP^+} \, , 
         \\
    \label{Eq:mixed-spin-TMDqbar-hT}
         h_T^{\bar q}(x,k_T) \; &=& \; 
         2P^+ \int dk^- \Biggl[
         \frac{x\,M^2 - P \cdot k}{M^2} \; A_{11}^{\bar q}
         \Biggr]_{k^+=xP^+} \, .
        \end{eqnarray}
\esub

\subsection{Polarized chiral even antiquark TMDs 
in pure-spin state case}

The polarized chiral even TMDs in the mixed-spin 
model are given by
\bsub
\label{Eq:TMDs-qbar-chiral-even-pure}
\begin{eqnarray}
    \label{Eq:pure-spin-TMDqbar-g1}
        g_1^{\bar q}(x,k_T) 
        &=& 
        2P^+ \int dk^- 
        \Biggl[
        \frac{x\,P \cdot k-x^2M^2-m_q^2}{M^2}\,A_8^{\bar q}
        -\frac{m_q}{M}\,x
            \biggl(\pm\,A_8^{\bar q}\biggr)
        \Biggr]_{k^+=xP^+}\,, 
        \\
    \label{Eq:pure-spin-TMDqbar-g1T}
         g_{1T}^{\perp \bar q}(x,k_T) \; &=& \; 
         2P^+ \int dk^-\Biggl[-x\,A_8^{\bar q}
         - \frac{m_q}{M}\,\biggl(\pm\,A_8^{\bar q}\biggr)\Biggr]_{k^+=xP^+}\,, 
         \\
    \label{Eq:pure-spin-TMDqbar-gT}
         g_T^{\bar q}(x,k_T) \; &=& \; 
         2P^+ \int dk^- 
         \Biggl[
         -\,\frac{\vec{k}_T^{\,2}+2m_q^2}{2M^2} \; A_8^{\bar q} 
         - \frac{m_q}{M}\;\frac{P\cdot k}{M^2}\,\biggl(\pm\,A_8^{\bar q}\biggr) 
         \Biggr]_{k^+=xP^+}\ \, , \end{eqnarray}
\esub
while $g^{\perp \bar q}_L(x,k_T)$ and 
$g^{\perp \bar q}_T(x,k_T)$ which do not depend
on the amplitude $A_{11}^{\bar q}$ are still given 
by the expressions 
(\ref{Eq:mixed-spin-TMDqbar-gLperp},~\ref{Eq:mixed-spin-TMDqbar-gTperp}).

\subsection{Polarized chiral odd antiquark TMDs 
in pure-spin state case }

The polarized chiral odd TMDs in the pure-spin state model
are given by
\bsub
\label{Eq:TMDs-qbar-chiral-odd-pure}
\begin{eqnarray}
    \label{Eq:pure-spin-TMDqbar-h1}
        h_1^{\bar q}(x,k_T) \; &=& \; 
        2P^+ \int dk^- \Biggl[
        \frac{\vec{k}_T^{\,2}-2\,x\,P\cdot k}{2M^2} \;
        \biggl(\pm\,A_8^{\bar q}\biggr) 
        - \frac{m_q}{M}\,x\,A_8^{\bar q}
        \Biggr]_{k^+=xP^+} \, ,\\
    \label{Eq:pure-spin-TMDqbar-h1Lperp}
         h^{\perp \bar q}_{1L}(x,k_T) \; &=& \; 
         2P^+ \int dk^- \Biggl[
         x\, \biggl(\pm\,A_8^{\bar q}\biggr) 
         +\frac{m_q}{M}\,A_8^{\bar q}
         \Biggr]_{k^+=xP^+} \, , 
\\
    \label{Eq:pure-spin-TMDqbar-h1Tperp}
         h^{\perp \bar q}_{1T}(x,k_T) \; &=& \; 
         2P^+ \int dk^- \Biggr[
         \biggl(\pm\,A_8^{\bar q}\biggr) 
         \Biggr]_{k^+=xP^+} \, , \\
    \label{Eq:pure-spin-TMDqbar-hL}
        h_L^{\bar q}(x,k_T) \; &=& \; 
        2P^+ \int dk^- 
        \Biggl[
        \frac{x^2M^2-2\,x\,P \cdot k}{M^2} 
        \biggl(\pm\,A_8^{\bar q}\biggr) 
        - \frac{m_q}{M}\,\frac{P \cdot k}{M^2}\;A_8^{\bar q}
         \Biggr]_{k^+=xP^+} \, , \\
    \label{Eq:pure-spin-TMDqbar-hTperp}
         h^{\perp \bar q}_T(x,k_T) \; &=& \; 
         2P^+ \int dk^- \Biggl[ -
         \frac{P\cdot k}{M^2}\,
         \biggl(\pm\,A_8^{\bar q}\biggr) 
         - \frac{m_q}{M}\,A_8^{\bar q}\Biggr]_{k^+=xP^+} 
         \, ,\\
    \label{Eq:pure-spin-TMDqbar-hT}
         h_T^{\bar q}(x,k_T) \; &=& \; 
         2P^+ \int dk^- \Biggl[ 
         \frac{x\,M^2-P \cdot k}{M^2} \; 
         \biggl(\pm\,A_8^{\bar q}\biggr) 
         \Biggr]_{k^+=xP^+} \, . 
\end{eqnarray}
\esub
Keeping in mind that quark mass effects are practically
negligible for the light $u$- and $d$-flavors, we see that the two $\pm$ solutions for 
$w^{\bar q}_{\pm}$ and $A_{\rm pol\pm}^{\bar q}$ in 
(\ref{Eq:qbar-solution-w2=1}--\ref{solutions-w-qbar-pm}),
predict opposite signs of the chiral odd polarized antiquark
TMDs. In order to determine the physical solution, we will
again rely on results from model and lattice calculations.

In order to determine the physical solution for 
$w^{\bar q}_{\pm}$ and $A_{\rm pol\pm}^{\bar q}$ in
(\ref{Eq:qbar-solution-w2=1}--\ref{solutions-w-qbar-pm})
we proceed analog to the quark case and consult 
theory results from literature. Far fewer studies are
available for antiquark distributions as compared to 
quarks, but they unanimously yield $h_1^{\bar q}$ and
$g_1^{\bar q}$ of opposite signs for a given flavor
${\bar q}$
\cite{
Pobylitsa:1996rs,Wakamatsu:1998rx,Schweitzer:2001sr,
Chen:2016utp,Kofler:2017uzq,Alexandrou:2021oih,Alexandrou:2018eet,Alexandrou:2021bbo,HadStruc:2021qdf}.
Now, the positive-sign solution $w^{\bar q}_+$ and 
$A_+^{\bar q}$ yields transversity and helicity 
antiquark TMDs of equal signs, i.e.\ this is the 
unphysical solution in antiquark case. 
The negative-sign solution 
$w^{\bar q}_-$ and $A_-^{\bar q}$ in
(\ref{Eq:qbar-solution-w2=1}--\ref{solutions-w-qbar-pm})
gives transversity and helicity antiquark TMDs of opposite
sign which is in agreement with results from literature.
We therefore choose the negative-sign solution in 
(\ref{Eq:qbar-solution-w2=1}--\ref{solutions-w-qbar-pm}) as the physical solution which is given by
\ba
    A_{11}^{\bar q} 
    &=& -\,A_8^{\bar q}\;, \phantom{\frac11} \nonumber  \\
    A_{\rm pol-}^{\bar q}  
    &=& \frac{P\cdot k - m_qM}{M^2}\;A_8^{\bar q},
    \nonumber\\
     w^\mu_{{\bar q}-}  &=& 
       S^\mu 
     - \frac{k\cdot S}{P\cdot k - m_qM} \; P^\mu 
     + \frac{M}{m_q}\,
     \frac{k\cdot S}{P\cdot k - m_qM} \; k^\mu\;.
     \label{solutions-qbar-final}
\ea
The final model expressions in the pure-spin state model
are given by
\bsub
\label{Eq:TMDs-qbar-final-pure}
\begin{eqnarray}
    \label{Eq:pure-final-TMDqbar-g1}
        g_1^{\bar q}(x,k_T) 
        &=& 
        2P^+ \int dk^- 
        \Biggl[
        \frac{x\,P \cdot k-x^2M^2+x\,m_qM-m_q^2}{M^2}\,A_8^{\bar q}
        \Biggr]_{k^+=xP^+}\,, 
        \\
    \label{Eq:pure-final-TMDqbar-g1T}
         g_{1T}^{\perp \bar q}(x,k_T) \; &=& \; 
         2P^+ \int dk^-\Biggl[
         -\,\frac{x\,M-m_q}{M}\; A_8^{\bar q}\Biggr]_{k^+=xP^+}\,,
        \\
    \label{Eq:pure-final-TMDqbar-gT}
         g_T^{\bar q}(x,k_T) \; &=& \; 
         2P^+ \int dk^- 
         \Biggl[
         -\,\frac{M\,\vec{k}_T^{\,2}-2\,m_q P\cdot k+2\,m_q^2M}{2M^3}
         \; A_8^{\bar q} \Biggr]_{k^+=xP^+}\ \, , \\
    \label{Eq:pure-final-TMDqbar-gLperp}
         g^{\perp \bar q}_L(x,k_T) \; &=& \;  
         2P^+ \int dk^- 
         \Biggl[\frac{P \cdot k-x\,M^2}{M^2}\; 
         A_8^{\bar q}\Biggr]_{k^+=xP^+}\ \, , 
       \\
    \label{Eq:pure-final-TMDqbar-gTperp}
         g^{\perp \bar q}_T(x,k_T) \; &=& \; 2P^+ \int dk^-
         \Biggl[-\,A_8^{\bar q}\;\Biggr]_{k^+=xP^+}\ \, ,\\
    \label{Eq:pure-final-TMDqbar-h1}
        h_1^{\bar q}(x,k_T) \; &=& \; 
        2P^+ \int dk^- \Biggl[
        \frac{2\,x\,P\cdot k-\vec{k}_T^{\,2}- 2\,x\,m_q M}{2M^2} \;
        A_8^{\bar q} \Biggr]_{k^+=xP^+} \, ,\\
    \label{Eq:pure-final-TMDqbar-h1Lperp}
         h^{\perp \bar q}_{1L}(x,k_T) \; &=& \; 
         2P^+ \int dk^- \Biggl[-\,
         \frac{x\,M-m_q}{M}\,A_8^{\bar q}
         \Biggr]_{k^+=xP^+} \, , 
         \\
    \label{Eq:pure-final-TMDqbar-h1Tperp}
         h^{\perp \bar q}_{1T}(x,k_T) \; &=& \; 
         2P^+ \int dk^- \Biggl[
         \,-\,A_8^{\bar q} \;
         \Biggr]_{k^+=xP^+} \, , \\
    \label{Eq:pure-final-TMDqbar-hL}
        h_L^{\bar q}(x,k_T) \; &=& \; 
        2P^+ \int dk^- 
        \Biggl[\;
        \frac{2\,x\,M\,P \cdot k-x^2M^3-m_q \, P \cdot k}{M^3} 
        \;A_8^{\bar q} \; \Biggr]_{k^+=xP^+} \, , \\
    \label{Eq:pure-final-TMDqbar-hTperp}
         h^{\perp \bar q}_T(x,k_T) \; &=& \; 
         2P^+ \int dk^- \Biggl[
         \frac{P\cdot k -m_q M}{M^2}\;A_8^{\bar q}\;
         \Biggr]_{k^+=xP^+} 
         \, ,  \\
    \label{Eq:pure-final-TMDqbar-hT}
         h_T^{\bar q}(x,k_T) \; &=& \; 
         2P^+ \int dk^- \Biggl[ 
         \frac{P \cdot k-M^2x}{M^2} \;A_8^{\bar q} 
         \Biggr]_{k^+=xP^+} \, .
\end{eqnarray}
\esub

\newpage
\section{Consistency of the approach}
\label{Sec-8:consistency}

In this section we demonstrate the internal
consistency of the approach. 

\subsection{\boldmath Sum rule of $e^a(x)$ }
\label{Subsec:e}

As a first consistency test we consider 
the sum rule for the twist-3 function $e^q(x)$ \cite{Jaffe:1991ra,Efremov:2002qh}.
Introducing the integrated functions
$e^a(x)=\int d^2k_T\,e^a(x,k_T)$ and
$f_1^a(x)=\int d^2k_T\,f_1^a(x,k_T)$
(we recall that in the parton model TMDs and
PDFs are simply related,
see Sec.~\ref{Subsec:def-TMD}), we obtain from
(\ref{Eq:rel-unp-q},~\ref{Eq:rel-unp-qbar}) the
result
\be\label{Eq:sum-e-1}
     \int_0^1 dx
     \biggl(x\,e^q(x) + x\,e^{\bar q}(x)\biggr)
     = \frac{m_q}{M}\int_0^1 dx
     \biggl(f_1^q(x) - f_1^{\bar q}(x)\biggr).
\ee
Using the customary "continuation" to negative $x$ according 
to $e^{\bar q}(x)=e^q(-x)$ and $f_1^{\bar q}(x)=-f_1^q(-x)$,
(\ref{Eq:sum-e-1}) can be expressed as
\be\label{Eq:sum-e-2}
    \int\limits_{-1}^1 dx\; x\,e^q(x)
    = \frac{m_q}{M}\,N_q \;,
\ee
where $N_q=\int_{-1}^1 dx f_1^q(x)$ is the number of valence
quarks of flavor $q$ in the nucleon. This is a consistency 
test for the amplitude relations derived from the equations 
of motion for respectively $\Psi_q$ and $\bar\Psi_q$
in Secs.~\ref{Sec-3:ampl-q} and \ref{Sec:ampl-qbar}.
The sum rule in (\ref{Eq:sum-e-2}) is correctly 
satisfied due to the opposite signs in the relations 
$A_1^q = \frac{m_q}{M}\,A_3^q$ and 
$A_1^{\bar q} = -\frac{m_q}{M}\,A_3^{\bar q}$.
We will follow up below on the negative-$x$ continuation 
also for the other TMDs. 

\subsection{Equation of motion (EOM) relations}
\label{Sec:eom-relations}

From the QCD equations of motion, one obtains the so-called 
EOM relations. Below we quote only the EOM relation for T-even
functions relevant in this work. The EOM relations among the 
unpolarized TMDs are given by
\bsub\label{Eq:EOMunp}
\ba \label{Eq:quark-case-unp-relations-1}
     x\,f^{\perp a}(x,k_T) = 
     x\,\tilde{f}^{\perp a}(x,k_T) + f_1^q(x,k_T), 
     \phantom{\frac11}\\
     \label{Eq:quark-case-unp-relations-2}
     x\,e^a(x,k_T) = 
     x\,\tilde{e}^a(x,k_T) + \frac{m_q}{M}\,f_1^a(x,k_T),
\ea\esub
where  $a=q,\;\bar q$. 
The EOM relations among the polarized TMDs read
\bsub\label{Eq:EOMpol}\ba
    xg_L^{\perp a}(x,k_T) 
    & = & x\tilde{g}_L^{\perp a}(x,k_T) 
    + 
    g_{1}^{a}(x,k_T) 
    + \frac{m_q}{M}\,h_{1L}^{\perp a}(x,k_T),
    \label{Eq:EOM-gLperp}\\
    xg_T^{a}(x,k_T)   	
    & = & x\tilde{g}_T^{a}(x,k_T) 
    +
    g_{1T}^{\perp(1)a}(x,k_T)
    + \frac{m_q}{M}\,h_1^{a}(x,k_T),
    \label{Eq:EOM-gT}\\
    xg_T^{\perp a}(x,k_T)
    & = & x\tilde{g}_T^{\perp a}(x,k_T) 
    + 
    g_{1T}^{\perp a}(x,k_T)
    + \frac{m_q}{M}\,h_{1T}^{\perp a}(x,k_T),
    \label{Eq:EOM-gTperp}\\
    xh_L^{a}(x,k_T)
    & = & x\tilde{h}_L^{a}(x,k_T) 
    - 
    2\,h_{1L}^{\perp(1)a}(x,k_T) 
    + \frac{m_q}{M}\,g_1^{a}(x,k_T),
    \label{Eq:EOM-hL}\\
    xh_T^{a}(x,k_T)            
    & = &  x\tilde{h}_T^{a}(x,k_T) 
    - 
    h_1^{a}(x,k_T) 
    - 
    h_{1T}^{\perp(1)a}(x,k_T)
    + \frac{m_q}{M}\,g_{1T}^{\perp a}(x,k_T),
    \label{Eq:EOM-hT}\\
    xh_T^{\perp a}(x,k_T)     
    & = &  x\tilde{h}_T^{\perp a}(x,k_T) 
    +
    h_1^{a}(x,k_T) 
    -
    h_{1T}^{\perp(1)a}(x,k_T),
    \phantom{\frac11} 
\label{Eq:EOM-hTperp}
\ea\esub
where the (1)-moments of TMDs are defined, 
for instance, as 
\be
    g_{1T}^{\perp(1) q}(x,k_T) 
    = \frac{\vec{k}_T^2}{2M^2}\,g_{1T}^{\perp q}(x,k_T)
\ee
and analogously for other TMDs. In QCD, the tilde terms 
in (\ref{Eq:EOMunp},~\ref{Eq:EOMpol}) are related to 
quark-gluon correlators \cite{Bacchetta:2006tn}. In quark 
models, despite the absence of gluonic degrees of freedom, 
the tilde terms are in general nonzero as they can arise 
from the respective model interactions. 

In the parton model, the tilde terms are expected to be
zero. One readily verifies that the model expressions 
for quark and antiquark TMDs in 
(\ref{Eq:rel-unp-q}, \ref{Eq:mixed-spin-TMDq-all},
\ref{Eq:pure-spin-TMDq-all}, \ref{Eq:rel-unp-qbar},
\ref{Eq:TMDs-qbar-chiral-even-mixed},
\ref{Eq:TMDs-qbar-chiral-odd-mixed},
\ref{Eq:TMDs-qbar-chiral-even-pure},
\ref{Eq:TMDs-qbar-chiral-odd-pure})
in mixed-spin and pure-spin version of the model 
satisfy the EOM relations with the tilde terms set to zero. 
This is another important consistency test of the approach.

\subsection{Relation between quark and antiquark correlators and its consequences}

In order to carry out further tests of the consistency 
of the approach, we explore the field-theoretical 
relation between the quark and antiquark correlators 
in (\ref{Eq:correlators}) given by
\cite{Tangerman:1994eh,Mulders:1995dh}
\be\label{Eq:Phi-q-qbar-relation}
    \Phi_{ij}^{\bar q}(k,P,S) 
    = - \, \Phi_{ij}^q(-k,P,S)
\ee
The relation in (\ref{Eq:Phi-q-qbar-relation}) 
shows that the amplitudes $A_i^q$ and $A_i^{\bar q}$ 
are related to each other by a "continuation" of 
$P\cdot k$ to $(-\,P\cdot k)$. From the expansions 
of the correlators $\Phi^a(k,P,S)$ in terms of the 
amplitudes in 
(\ref{Eq:correlator-decompose}) and the relation 
in (\ref{Eq:Phi-q-qbar-relation}) we read off
\be\label{Eq:ampl-q-qbar-relations}
     A_i^{\bar q}(P\cdot k) = 
     \begin{cases}
        + A_i^q(-P\cdot k)
        & {\rm for} \quad i = 3,\, 4,\, 5,\, 7,\,10, \, 12,
        \cr
        - A_i^q(-P\cdot k) 
        & {\rm for} \quad i = 1,\,2,\, 6,\,8,\,9,\, 11. 
        \end{cases}
\ee
As a first consistency test, let us rederive 
the relations among the antiquark amplitudes in 
(\ref{Eq:ampl-01-bar},~\ref{Eq:ampl-02-bar})
from the relations among the quark amplitudes 
$A^q(P\cdot k)$ in 
(\ref{Eq:ampl-02},~\ref{Eq:ampl-04}). 
This is accomplished by continuing $k \to (-k)$ in
(\ref{Eq:ampl-02},~\ref{Eq:ampl-04}).
Using (\ref{Eq:ampl-q-qbar-relations}) to express the
$A^q_i(-P\cdot k)$ in terms of the $A^{\bar q}_i(P\cdot k)$
yields the antiquark amplitude relations in 
(\ref{Eq:ampl-01-bar},~\ref{Eq:ampl-02-bar}).
This test demonstrates the consistency of the derivations 
of the relations (\ref{Eq:ampl-02},~\ref{Eq:ampl-04}) 
among the quark amplitudes and the  derivations of the 
relations (\ref{Eq:ampl-01-bar},~\ref{Eq:ampl-02-bar})
among antiquark amplitudes.

As a second test related to the amplitudes, let us
consider the final results for the correlators in the
pure-spin state model with massive partons which are given by
\bsub\label{Eq:fin-correlator-all}
\ba\label{Eq:fin-correlator}
    \Phi^q(k,P,S) 
    &=& 
    (\slashed{k}+m_q)\;\biggl[
    A_{\rm unp}^q(P\cdot k)
    + \gamma_5\,\slashed{w}_q(k)\,A_{\rm pol}^q(P\cdot k)
    \biggr]\\
    \Phi^{\bar q}(k,P,S) 
    &=&
    (\slashed{k}-m_q)\,\biggl[
    A_{\rm unp}^{\bar q}(P\cdot k) 
    + \gamma_5\,\slashed{w}_{\bar q}(k)
    \,A_{\rm pol}^{\bar q}(P\cdot k)\biggr] \\
    A_{\rm unp}^a(P\cdot k) 
    &=& A_3^a(P\cdot k)\,, \phantom{\frac11}\\
    A_{\rm pol}^q(P\cdot k)
    &=& 
    \frac{P\cdot k+m_qM}{M^2}\,A_8^q(P\cdot k)\,,\\
    A_{\rm pol}^{\bar q}(P\cdot k) 
    &=& 
    \frac{P\cdot k-m_qM}{M^2}\;A_8^{\bar q}(P\cdot k)\,,\\
    w_q^\mu(k)  
    &=& 
       S^\mu 
     - \frac{k\cdot S}{P\cdot k+m_qM}\;P^\mu 
     - \frac{M}{m_q}\,
     \frac{k\cdot S}{P\cdot k+m_qM}\;k^\mu  \, ,\\
     w^\mu_{\bar q}(k)
     &=& 
       S^\mu 
     - \frac{k\cdot S}{P\cdot k-m_qM} \;P^\mu 
     + \frac{M}{m_q}\,
     \frac{k\cdot S}{P\cdot k-m_qM} \; k^\mu = w_q^\mu(-k) 
     \;.
\ea\esub
From (\ref{Eq:ampl-q-qbar-relations},
\ref{Eq:fin-correlator-all}) we see that
$A_{\rm unp}^{\bar q}(P\cdot k)=A_{\rm unp}^q(-P\cdot k)$ and
$A_{\rm pol}^{\bar q}(P\cdot k)=A_{\rm pol}^q(-P\cdot k)$
while $w^\mu_{\bar q}(k)=w_q^\mu(-k)$. With these preparations,
we see that the quark and antiquark correlators are related to each 
other by the field-theoretical relation 
in (\ref{Eq:Phi-q-qbar-relation}).
This test is non-trivial, because we have independently 
chosen the signs of the solutions in the quark and antiquark cases in 
(\ref{Eq:q-solutions-options},~\ref{Eq:qbar-solutions-options}).
In the same way, one can show that the quark and antiquark
correlators are consistently described also in the 
mixed-spin state and massless versions of the model.

\subsection{Continuation of quark TMDs to negative $x$}
\label{Subsec:cont-neg-x}

The model expression for $f_1^q(x,k_T)$ in
(\ref{Eq:f1-q-final}) can be expressed as
\bsub\be\label{Eq:showing-cont-neg-x-f1}
    f_1^q(x,k_T) 
    = 2P^+\int dk^-\int dk^+\,\delta(k^+ - xP^+)
      \biggl[xA_3^q(P\cdot k)\biggr] \,.
\ee   
In this expression, we replace $x$ by $(-x)$, and
perform the substitutions\footnote{Notice that
    under the substitutions $k^\pm \to -k^\pm$ in 
    the integrals in (\ref{Eq:showing-cont-neg-x-f1}),
    the product $P\cdot k$ changes sign
    $P\cdot k \to -P\cdot k$. This is so because 
    $P\cdot k = P^+k^- + P^-k^+$ 
    is independent of $k_T$ as by definition 
    the  nucleon momentum has no transverse component.
    \label{footnote-P.k}}
$k^\pm \to -k^\pm$. This yields
\ba
      f_1^q(-x,k_T) 
      = 
      -\,2P^+\int dk^-\int dk^+\,\delta(k^+ - xP^+)
      \biggl[xA_3^q(-P\cdot k)\biggr] \nonumber && \\
      = 
      -\,2P^+\int dk^-\int dk^+\,\delta(k^+ - xP^+)
      \biggl[xA_3^{\bar q}(P\cdot k)\biggr]   
      \label{Eq:deriving-f1-conitinued-negative-x}
      &&
\ea
\esub
where in the last step we made use of
(\ref{Eq:ampl-q-qbar-relations}). 
Comparing to (\ref{Eq:TMDqbar-f1}) 
we recognize the expression for
$(- 1)\, f_1^{\bar q}(x,k_T)$.
We can proceed 
analogously with the other TMDs. Summarizing
the results, we find the familiar relations
\bsub\label{Eq:relations-TMDs-q-qbar}
\begin{alignat}{2}
    f_1^{\bar q}(x,k_T)         = & - &  f_1^q(-x,k_T) \,,\\
    g_1^{\bar q}(x,k_T)         = &   &  g_1^q(-x,k_T) \,,\\
    h_1^{\bar q}(x,k_T)         = & - &  h_1^q(-x,k_T) \,,\\
    g_{1T}^{\perp\bar q}(x,k_T) = & - &  g_{1T}^{\perp q}(-x,k_T) \,,\\
    h_{1L}^{\perp\bar q}(x,k_T) = &  &  h_{1L}^{\perp q}(-x,k_T) \,,\\
    h_{1T}^{\perp\bar q}(x,k_T) = & - &  h_{1T}^{\perp q}(-x,k_T) \,,\\
    e^{\bar q}(x,k_T)           = &   &  e^q(-x,k_T) \,,\\
    f^{\perp\bar q}(x,k_T)      = &   &  f^{\perp q}(-x,k_T) \,,\\
    g_T^{\bar q}(x,k_T)         = &   &  g_T^q(-x,k_T) \,,\\
    g_T^{\perp\bar q}(x,k_T)    = &   &  g_T^{\perp q}(-x,k_T) \,,\\
    g_L^{\perp\bar q}(x,k_T)    = & - &  g_L^{\perp q}(-x,k_T) \,,\\
    h_L^{\bar q}(x,k_T)         = & - &  h_L^q(-x,k_T) \,,\\
    h_T^{\perp\bar q}(x,k_T)    = &   &  h_T^{\perp q}(-x,k_T) \,,\\
    h_T^{\bar q}(x,k_T)         = &   &  h_T^q(-x,k_T) 
    \,.
\end{alignat}
\esub
In these relations $x$ is always in the range
$0< x < 1$. We remark that based on these relations 
it is customary to introduce TMDs defined on the 
domain $-1 < x < 1$ with the understanding that 
quark TMD functions of $(-x)$ mean $(\pm 1)$ of the 
respective antiquark TMD functions of $x$ with the 
signs as specified in (\ref{Eq:relations-TMDs-q-qbar}), 
e.g., $h_1^q(x,k_T)$ for  $0<(-x)<1$ means 
$-\,h_1^{\bar q}(x,k_T)$ for $0<x<1$, etc,
cf.\ also Sec.~\ref{Subsec:e}.

\section{Evaluation of TMDs in Covariant Parton Model}
\label{Sec-9:evaluation-in-CMP}

In this section we will show that the pure-spin state parton model derived in this work corresponds to the CPM of Refs.~\cite{Zavada:1996kp,Zavada:2001bq,
Zavada:2002uz,Efremov:2004tz,Zavada:2007ww,Efremov:2009ze,
Zavada:2009ska,Efremov:2009vb,Efremov:2010mt,Zavada:2011cv,
Zavada:2013ola,Zavada:2015gaa,Zavada:2019yom,Bastami:2020rxn}.
For that we will introduce the notation of \cite{Zavada:1996kp,Zavada:2001bq, Zavada:2002uz}, explore consequences of the equation of motion \cite{DAlesio:2009cps}, consider kinematic DIS constraints, and rederive the model expressions for quark TMDs from prior studies \cite{Efremov:2009ze,Efremov:2010mt,Bastami:2020rxn}
and present new results for antiquark TMDs.

\subsection{Model expressions}

Starting from the equation of motion $(i\slashed{\partial}-m_q)\,\Psi^q(z)=0$, one obtains for the quark correlator the relation \cite{DAlesio:2009cps},
\ba
	0 
	&=& 
	\int \frac{\mathrm{d}^4z}{(2\pi)^4}\;\mathrm{e}^{i k z}\,
    \langle N|\,\overline{\Psi}_j^{\,q}(0)\;
    \biggl[\,
    (i\overrightarrow{\slashed{\partial}}+m_q)_{ik}^{ }
    (i\overrightarrow{\slashed{\partial}}-m_q)_{kl}^{ }
    \Psi_l^q(z)\,\biggr]|N\rangle \nonumber\\
    &=&
    \int \frac{\mathrm{d}^4z}{(2\pi)^4}\;\mathrm{e}^{i k z}\,
    \langle N|\,\overline{\Psi}_j^{\,q}(0)\;
    \biggl[\,(\,-\,\square - m_q^2\,)\,\mathbb{1}_{il}^{ }
    \Psi_l^q(z)\biggr]
    |N\rangle
    \,, \ \nonumber\\
    &=& 
    (k^2-m^2)\,\phi^q_{ij}(k,P,S)\,, \phantom{\int} \label{Eq:eval-01}
\ea
where in the intermediate step we have performed twice
integration by parts. The antiquark correlator satisfies 
an analogous relation. Since the Lorentz structures in the
decompositions of the correlators are linearly independent,
the result (\ref{Eq:eval-01}) and the analogous result 
for the antiquark correlator imply that the amplitudes
satisfy
\be\label{Eq:onshell-0}
    (k^2-m^2)\,A_i^a(P\cdot k,\,k^2) = 0,
    \quad a = q,\;\bar{q}\,.
\ee
Keeping in mind that in the pure-spin-state version of the model we only have 2 independent amplitudes which can be expressed as $A_{\rm unp}^a(P\cdot k,k^2)$ and $A_{\rm pol}^a(P\cdot k,k^2)$, cf.\ (\ref{Eq:fin-correlator-all}), the solutions
to (\ref{Eq:onshell-0}) can be stated as 
\ba\label{Eq:onshell-1}
   A_{\rm unp}^a(P\cdot k,k^2) =  M\,\delta(k^2-m_q^2)\,\Theta_{\rm kin}(P\cdot k)\,
   {\cal G}^a(P\cdot k) \,,\nonumber\\
   A_{\rm pol}^a(P\cdot k,k^2) =  M\,\delta(k^2-m_q^2)\,\Theta_{\rm kin}(P\cdot k)\,
   {\cal H}^a(P\cdot k)\,,
\ea
where $a=q,\;\bar q$.
Several comments are in order. The delta-function
$\delta(k^2-m_q^2)$ ensures that (\ref{Eq:onshell-0}) 
is satisfied, and puts the partons on shell. The functions
${\cal G}^q(P\cdot k)$ and ${\cal H}^q(P\cdot k)$ are
Lorentz-invariant functions of the variable $P\cdot k$ 
and are defined following the notation of
\cite{Zavada:1996kp,Zavada:2001bq,Zavada:2002uz}.
In (\ref{Eq:onshell-1}), the factor $M$ was 
introduced for convenience such that 
${\cal G}^q(P\cdot k)$ and ${\cal H}^q(P\cdot k)$ have the
dimension (mass)$^{-3}$ and can be interpreted as 3D momentum
densities \cite{Zavada:1996kp,Zavada:2001bq,Zavada:2002uz}.
Finally, the function $\Theta_{\rm kin}(P\cdot k)$
incorporates the kinematic constraints on the partons 
and is defined as
\be\label{Eq:onshell-1a}
    \Theta_{\rm kin}(P\cdot k) 
    = \Theta_{\rm kin}^+(P\cdot k)
    + \Theta_{\rm kin}^-(P\cdot k),
    \quad
    \Theta_{\rm kin}^\pm(P\cdot k)
    = \Theta(\pm P\cdot k)\;
    \Theta\left((P\mp k)^2\right)\,.
\ee
Below we shall see that only $\Theta_{\rm kin}^+(P\cdot k)$
in (\ref{Eq:onshell-1a}) contributes to quark 
or antiquark TMDs. However, although it drops out from 
 TMDs, the presence of $\Theta_{\rm kin}^-(P\cdot k)$ in 
(\ref{Eq:onshell-1a}) is nevertheless of importance
for the analytic structure of the amplitudes and 
completeness of the model. We will follow up on 
this shortly. Let us mention here merely that due to
$\Theta(P\cdot k) = \Theta(k^0)$, the role of $\Theta(P\cdot k)$ in $\Theta_{\rm kin}^+(P\cdot k)$ 
is to project out positive energy solutions
\cite{Bastami:2020rxn}, while $\Theta((P-k)^2)$ 
ensures that the nucleon remnant has
positive energy and constitutes a physical state 
\cite{DAlesio:2009cps,Ellis:1982cd}.

The scalar product $P\cdot k$ is positive (since it 
can be evaluated in any frame including nucleon rest 
frame where it is $Mk^0$; and for a real, on-shell 
parton the energy $k^0$ is of course positive). 
But it is convenient to define the 
``analytical continuation'' of the covariant functions 
at negative values of $P\cdot k$ as follows
\bsub\be\label{Eq:onshell-1b}
    {\cal G}^q(-P\cdot k) =
    {\cal G}^{\bar q}(P\cdot k), \quad
    {\cal H}^q(-P\cdot k) =
    {\cal H}^{\bar q}(P\cdot k).
\ee
With this definition, we see that the amplitudes defined in (\ref{Eq:onshell-1}) satisfy
\be\label{Eq:Aq-Aqbar-relation-model}
    A_{\rm unp}^q(-P\cdot k,k^2) =
    A_{\rm unp}^{\bar q}(P\cdot k,k^2)\,, \quad
    A_{\rm pol}^q(-P\cdot k,k^2) =
    A_{\rm pol}^{\bar q}(P\cdot k,k^2)\,.
\ee
We also see that with these definitions the model expressions for the quark and anti-quark correlators are given by 

\ba\label{Eq:corr-q-contact-to-previous-notation}
    \Phi^q(k,P,S)
    &=& 
    (\slashed{k}+m_q)\biggl({\cal G}^q(P\cdot k)
    +{\cal H}^q(P\cdot k)\,\gamma_5\,\slashed{w}{ }_q(k)\biggr)\,
    M\,\delta(k^2-m_q^2)\,\Theta_{\rm kin}(P\cdot k) \\
     \Phi^{\bar q}(k,P,S) &=& 
     (\slashed{k}-m_q)\biggl({\cal G}^{\bar q}(P\cdot k)
    +{\cal H}^{\bar q}(P\cdot k)\,\gamma_5\,\slashed{w}{ }_{\bar q}(k)\biggr)\,
    M\,\delta(k^2-m_q^2)\,\Theta_{\rm kin}(P\cdot k) 
\ea\esub
with $w_a^\mu(k)$ defined in
(\ref{Eq:fin-correlator-all}), and satisfy
(\ref{Eq:Phi-q-qbar-relation}). This is
important for the internal consistency of the
model. Notice that a consistent description of 
the quark and antiquark correlators is guaranteed by 
the specific structure of $\Theta_{\rm kin}(P\cdot k)$
as defined in (\ref{Eq:onshell-1a}). 

In the quark case, the result in
(\ref{Eq:corr-q-contact-to-previous-notation})
coincides with the expression for the quark correlator 
in Ref.~\cite{Bastami:2020rxn}.\footnote{%
    At this occasion, let us remark that in
    Ref.~\cite{Bastami:2020rxn} instead of 
    nucleon mass $M$ the nucleon energy $P^0$ 
    was used to give ${\cal G}^q(P\cdot k)$ 
    and ${\cal H}^q(P\cdot k)$ the desired dimension. 
    This choice is incorrect as it would 
    imply incorrect Lorentz-transformation 
    properties for the amplitudes. 
    But in \cite{Bastami:2020rxn} the TMD 
    expressions were evaluated in the nucleon rest 
    frame where $P^0=M$, so the practical results from
    \cite{Bastami:2020rxn} are correct.}
Notice that the step function
$\Theta((P-k)^2)$ was implicitly understood in 
Refs.~\cite{Zavada:1996kp,Zavada:2001bq,
Zavada:2002uz,Efremov:2004tz,Zavada:2007ww,Efremov:2009ze,
Zavada:2009ska,Efremov:2009vb,Efremov:2010mt,Zavada:2011cv,
Zavada:2013ola,Zavada:2015gaa,Zavada:2019yom,Bastami:2020rxn}
(and explicitly formulated in
\cite{DAlesio:2009cps}).
With the above remarks in mind, the result in
(\ref{Eq:corr-q-contact-to-previous-notation}) 
practically reproduces the results for quark TMDs 
from \cite{Bastami:2020rxn}. 
In the antiquark case, the result in
(\ref{Eq:corr-q-contact-to-previous-notation}) 
is new and was not given in prior studies.

Let us explicitly carry out the calculation 
of the unpolarized leading-twist quark TMD. 
Starting from (\ref{Eq:f1-q-final}) and 
inserting for 
$A_3^q(P\cdot k,k^2)=A_{\rm unp}^q(P\cdot k,k^2)$ 
the result in (\ref{Eq:onshell-1}) we obtain 
\bsub\ba\label{Eq:eval-f1-parts}
    f_1^q(x,k_T) 
    =
    2P^+ \int dk^- \int dk^+\,\delta(k^+-xP^+)\,
    x M\,\delta(k^2-m_q^2)\,\Theta_{\rm kin}(P\cdot k)\,
    {\cal G}^q(P\cdot k) 
    = (\mbox{part 1}) + (\mbox{part 2})
\ea
with the two parts arising from respectively the two
contributions $\Theta_{\rm kin}^\pm(P\cdot k)$
in (\ref{Eq:onshell-1a}). 
The first part in (\ref{Eq:eval-f1-parts}) 
is evaluated by noticing that
$\Theta(P\cdot k)=\Theta(k^0)$ and using
$\delta(k^2-m_q^2)\,\Theta(k^0)=\delta(k^0-E_q)\,/\,(2E_q)$
with 
\be\label{Eq:define-Eq}
    E_q=\sqrt{\vec{k}{ }^2+m_q^2}
\ee
denoting the parton energy. Recalling that
$k^\pm=(k^0\pm k^1)/\sqrt{2}$, changing the integration
variables $dk^+dk^- \to dk^0 dk^1$ and choosing for
convenience to work in the nucleon rest frame, we obtain 
\ba
    (\mbox{part 1}) 
    &=& xM\int dk^0 \int dk^1\:
    \delta\biggl(x-\frac{k^+}{P^+}\biggr)\,
    \frac{\delta(k^0-E_q)}{E_q}\:
    \Theta_{\rm kin}^+(P\cdot k)\:
    {\cal G}^q(P\cdot k)
    \label{Eq:eval-f1-part-1A}\\
    &=&
    x\,M \int\frac{dk^1}{E_q}\;
    \delta\biggl(x-\frac{E_q+k^1}{M}\biggr)\;
    \Theta\left(M^2+m_q^2-2ME_q\right)\;
    {\cal G}^q(ME_q)\,.
    \label{Eq:eval-f1-part-1B}
\ea

The ``part 2'' is evaluated similarly except 
that now we pick $\Theta(-P\cdot k) = \Theta(-k^0)$ from 
$\Theta_{\rm kin}(P\cdot k)$ in (\ref{Eq:onshell-1a}). 
We then get 
$\delta(k^2-m_q^2)\,\Theta(-k^0) = \delta(k^0+E_q)\,/\,(2E_q)$ 
which yields, using the nucleon rest frame for convenience,
the result
\ba\label{Eq:eval-f1-part-2}
    (\mbox{part 2}) = 
    xM \int \frac{dk^1}{E_q}\,\delta\biggl(x+\frac{E_q-k^1}{M}\biggr)
    \Theta(M^2+m_q^2+2ME_q)\,{\cal G}^q(-ME_q) 
    = 0\;.
\ea\esub
The ``part 2'' vanishes because $x$ is positive 
and $(E_q-k^1)/M$ is also always positive for an 
on-shell particle such that the delta-function under 
the $k^1$-integral in (\ref{Eq:eval-f1-part-2})
is always zero.

The result for $f_1^q(x,k_T)$ due to ``part 1'' in
(\ref{Eq:eval-f1-parts}) coincides exactly with 
the expression from prior studies 
\cite{Zavada:1996kp,Zavada:2001bq,
Zavada:2002uz,Efremov:2004tz,Zavada:2007ww,Efremov:2009ze,
Zavada:2009ska,Efremov:2009vb,Efremov:2010mt,Zavada:2011cv,
Zavada:2013ola,Zavada:2015gaa,Zavada:2019yom,Bastami:2020rxn}
(notice that in prior studies $ {\cal G}^q(P\cdot k)$ 
was often denoted in the nucleon rest frame as 
${\cal G}^q(E_q)$ or ${\cal G}^q(k^0)$ for simplicity).
The calculation of $f_1^{\bar q}(x,k_T)$ is analog 
to the above calculation of the quark TMD with the 
labels $q \leftrightarrow \bar q$ interchanged, and 
other quark and antiquark TMDs are evaluated
analogously. In order to list the final results and
abbreviate the model expressions, it is convenient to 
introduce a compact notation for the flavor-dependent 
integration measures \cite{Efremov:2009ze}
\bsub\label{Eq:compact-measure}
\ba
    \{dk^1\}^a_{\rm unp}
    &=&
    \frac{dk^1}{E_q}\;\frac{{\cal G}^a(ME_q)}{E_q+m_q}\;
    \delta\biggl(x-\dfrac{E_q+k^1}{M}\biggr)\;
    \Theta(M^2+m_q^2-2ME_q)\, , \\
    \{dk^1\}^a_{\rm pol}
    &=&
    \frac{dk^1}{E_q}\;\frac{{\cal H}^a(ME_q)}{E_q+m_q}\;
    \delta\biggl(x-\dfrac{E_q+k^1}{M}\biggr)\;
    \Theta(M^2+m_q^2-2ME_q)\, ,
\ea
\esub
Summarizing all results in the compact notation of 
(\ref{Eq:compact-measure}), we obtain
\bsub
\label{Eq:final-results-TMDs}
\begin{eqnarray}
    f_1^a(x,k_T)&=&\int\{dk^1\}^a_{\rm unp}
    \biggl[xM(E_q+m_q)\biggr]
    ,\\
    g_1^a(x,k_T)&=&\int\{dk^1\}^a_{\rm pol}\, \biggl[x^2M^2-xE_qM+xm_qM+m_q^2\biggr]
    ,\\
    g_{1T}^{\perp a}(x,k_T)&=&\int\{dk^1\}^a_{\rm pol}\, 
    \biggl[xM^2+m_qM\biggr]
    ,\\
    h_{1}^{a}(x,k_T)&=&\int\{dk^1\}^a_{\rm pol}\, 
    \biggl[2xE_qM-\vec{k}_T^2+2xm_qM\biggr]
    ,\\
    h_{1L}^{ \perp a}(x,k_T)&=&\int\{dk^1\}^a_{\rm pol}\,
    \biggl[-xM^2-m_qM \biggr]
    ,\\
    h_{1T}^{ \perp a}(x,k_T)&=&\int\{dk^1\}^a_{\rm pol}\, 
    \biggl[- M^2  \biggr] 
    ,\\
    e^{a}(x,k_T)&=&\int\{dk^1\}^a_{\rm unp}
    \biggl[ m_q(E_q+m_q) \biggr]
    ,\\
    f^{ \perp a}(x,k_T)&=&\int\{dk^1\}^a_{\rm unp}
    \biggl[ M(E_q+m_q) \biggr]
    ,\\
    g_{T}^{ a}(x,k_T)&=&\int\{dk^1\}^a_{\rm pol}\, 
    \biggl[\vec{k}_T^2+2m_qE_q+2m_q^2 \biggr]
    ,\\
    g_{L}^{\perp a}(x,k_T)&=&\int\{dk^1\}^a_{\rm pol}\, 
    \biggl[xM^2-E_q M  \biggr]
    ,\\
    g_{T}^{\perp a}(x,k_T)&=&\int\{dk^1\}^a_{\rm pol}\, 
    \biggl[ M^2 \biggr]
    ,\\
    h_{L}^{  a}(x,k_T)&=&\int\{dk^1\}^a_{\rm pol}\, 
    \biggl[2xME_q-x^2M^2+m_qE_q \biggr]
    ,\\
    h_{T}^{ a}(x,k_T)&=&\int\{dk^1\}^a_{\rm pol}\, 
    \biggl[ E_qM-xM^2 \biggr],\\
    h_{T}^{\perp  a}(x,k_T)&=&\int\{dk^1\}^a_{\rm pol}\, 
    \biggl[ M(E_q+m_q)  \biggr]
    .
\end{eqnarray}
\esub
These results are valid for $a=q,\,\bar q$. The results for
quark TMDs were obtained before in \cite{Efremov:2009ze}
in twist-2 and in \cite{Bastami:2020rxn} in twist-3 case.
The results for antiquark TMDs are presented for the
first time in this work. 
It should be noted that equivalent expressions can
be obtained for the TMDs  \cite{Efremov:2009ze}
by exploring  
$E_q^2 = k_1^2 + \vec{k}_T^2 + m_q^2$ and $E_q+k^1=xM$
due to the on-shell condition and the delta-function
present in the compact definition of the integration
measure (\ref{Eq:compact-measure}).

The model expression for quark and antiquark TMDs satisfy 
(\ref{Eq:relations-TMDs-q-qbar}).
This follows immediately from the properties of the
amplitudes in (\ref{Eq:Aq-Aqbar-relation-model}).
It can also be verified explicitly by working
with the model expressions at the level of (\ref{Eq:eval-f1-part-1A})
before the $k^0$-integration is carried out and 
the contribution of ``part~2'' explicitly drops out. 
In fact, when we continue the expression for
the quark TMD to negative $x$ values, then the roles 
of $\Theta_{\rm kin}^+$ and $\Theta_{\rm kin}^-$ 
in (\ref{Eq:onshell-1a}) interchange: 
when evaluating $f_1^q(-x,k_T)$ the contribution
from $\Theta_{\rm kin}^+$ drops out and that of
$\Theta_{\rm kin}^-$ yields a non-zero result.
In order to show that 
$f_1^q(-x,k_T)=-f_1^{\bar q}(x,k_T)$ 
one needs to perform substitutions $k^0\to -k^0$
and $k^1\to -k^1$ (or equivalently
$k^\pm\to-k^\pm$ in lightcone coordinates).
Under these substitutions $P\cdot k\to -P\cdot k$,
cf.\ footnote~\ref{footnote-P.k}, such 
that $\Theta_{\rm kin}^+(P\cdot k)$ is transformed into
$\Theta_{\rm kin}^-(P\cdot k)$ and vice versa.
Thus, we see that the full analytical structure 
of $\Theta_{\rm kin}(P\cdot k)$ 
is crucial for the proof of the relations
in (\ref{Eq:relations-TMDs-q-qbar}).
Once the integration over $k^0$ is carried 
out and the ``part 2'' contribution drops out 
in (\ref{Eq:eval-f1-part-2}), the relations 
(\ref{Eq:relations-TMDs-q-qbar}) are implicit
and cannot be easily verified.

\subsection{Relations among antiquark TMDs in the CPM}
\label{Secsec:CQM-qbar-relations}

In QCD all TMDs are independent functions. But simpler
model dynamics or additional model symmetries can 
generate relations between TMDs in models. 
In the quark case, relations among TMDs were
found in several models. 
Based on the results of the previous section we present
relations among antiquark TMDs which to the best of our
knowledge have not been derived before.
In fact, as shown in the previous section,
the antiquark TMDs are formally given by same
expressions as the quark TMDs with 
${\cal G}^q(P\cdot k)$ replaced by 
${\cal G}^{\bar q}(P\cdot k)$ and analog for
${\cal H}^{\bar q}(P\cdot k)$. Therefore, 
the antiquark TMDs and quark TMDs satisfy 
in the CPM the same model relations.
Considering how~little is known from models about 
nonperturbative properties of antiquark TMDs,
the new relations are of interest.

In Sec.~\ref{Sec:eom-relations} we already discussed 
the EOM relations (\ref{Eq:EOMunp},~\ref{Eq:EOMpol}), 
which are satisfied in the model with the tilde terms
absent also in the antiquark case. 
Next, we consider the so-called 
quark model Lorentz invariance relations (qLIRs) which
arise in effective theories without gluonic degrees
of freedom where T-odd $A_i^a$ amplitudes are absent
and the 14 T-even TMDs are described in terms of 9
T-even amplitudes implying 5 relations given for 
antiquarks by
\cite{Mulders:1995dh}
\bsub\label{Eq:LIRs}
\begin{align}
   \label{eq:LIR1} g_T^{\bar q}(x) 
   \; \stackrel{\rm qLIR}{=}
   & \; g_1^{\bar q}(x) + \frac{\mathrm{d}}{\mathrm{d} x} g^{\perp(1){\bar q}}_{1T}(x)\, ,\\ 
   \label{eq:LIR2} h_L^{\bar q}(x) 
   \; \stackrel{\rm qLIR}{=}
   & \; h_1^{\bar q}(x) - \frac{\mathrm{d}}{\mathrm{d} x} h^{\perp(1){\bar q}}_{1L}(x) \, , \\
   \label{eq:LIR3} h_T^{\bar q}(x) 
   \;  \stackrel{\rm qLIR}{=}
   & \; - \frac{\mathrm{d}}{\mathrm{d} x} 
   h^{\perp(1){\bar q}}_{1T}(x) \, , \\
   \label{eq:LIR4} g_L^{\perp \bar q}(x) + \frac{\mathrm{d}}{\mathrm{d} x} 
   g_T^{\perp(1)\bar q}(x)   
   \;\stackrel{\rm qLIR}{=}& \; 0 \, ,\\ 
   \label{eq:LIR5}
   h_T^{\bar q}(x,p_T)-
   h_T^{\perp \bar q}(x,p_T) 
   \; \stackrel{\rm qLIR}{=}
   & \; h^{\perp \bar q}_{1L}(x,p_T) \, .
\end{align}
\esub
The qLIRs are written such that twist-3 TMDs 
appear on the left-hand sides and twist-2 TMDs
(if any) on the right-hand-sides. The CPM model
expressions for antiquark TMDs 
(\ref{Eq:final-results-TMDs})
satisfy the qLIRs (\ref{Eq:LIRs}).
The proofs are identical to the proofs of the
corresponding qLIRs for quark TMDs and can be 
found in Ref.~\cite{Efremov:2009ze}.

While the qLIRs (\ref{Eq:LIRs}) must be valid
in all models with no explicit gluon degrees of
freedom, in specific models further model relations
may hold. In the CPM, the antiquark TMDs obey
the following model-specific relations
\bsub\label{Eq:quark-model-relations}
\ba
      g_{1T}^{\perp \bar q}(x,p_T) 
      &=& - h_{1L}^{\perp \bar q}(x,p_T), 
      \label{Eq:qm-rel-1}\\
      g_{ T}^{\perp \bar q}(x,p_T) 
      &=& - h_{1T}^{\perp \bar q}(x,p_T), 
      \label{Eq:qm-rel-2}\\
      g_{ L}^{\perp \bar q}(x,p_T) 
      &=& - h_{ T}^{\bar q}(x,p_T), 
      \label{Eq:qm-rel-3}\\
      g_1^{\bar q}(x,p_T)-h_1^{\bar q}(x,p_T) 
      &=& h_{1T}^{\perp(1)\bar q}(x,p_T), 
      \label{Eq:qm-rel-4}\\
      g_T^{\bar q}(x,p_T)-h_L^{\bar q}(x,p_T) 
      &=& h_{1T}^{\perp(1)\bar q}(x,p_T), 
      \label{Eq:qm-rel-5}
\ea
and one more relation which coincides with the qLIR
(\ref{eq:LIR5}). The quark-model relations are
verified by directly inserting the
model expressions (\ref{Eq:final-results-TMDs}).
The analogous relations among quark TMDs 
were derived in CPM in \cite{Avakian:2010br} 
and are valid also in spectator, bag, and
light-front constituent quark model 
\cite{Jakob:1997wg,Avakian:2009jt,Avakian:2010br,Pasquini:2008ax,Lorce:2011zta}.
The CPM also supports the following non-linear relations among antiquark TMDs
\ba
    \frac12\biggl[h_{1L}^{\perp \bar q}(x,p_T)\biggr]^2
    &=&     
    -\,h_1^{\bar q}(x,p_T)\,h_{1T}^{\perp \bar q}(x,p_T)
    \,,
    \label{Eq:qm-rel-7}\\
    \frac12\biggl[g_{1T}^{\perp \bar q}(x,p_T)\biggr]^2
    &=&
    g_{1T}^{\perp\bar q}(x,p_T)\,g_{L}^{\perp\bar q}(x,p_T)+
    g_T^{\bar q}(x,p_T)\,g_{T}^{\perp\bar q}(x,p_T)
    \,.
    \label{Eq:qm-rel-8}
\ea
\esub
The linear and nonlinear relations were 
derived for quark TMDs in \cite{Efremov:2009ze}
in twist-2 and in \cite{Bastami:2020rxn}
in twist-3 case.

For twist-2 quark TMDs, the deeper reason 
underlying the model relations
(\ref{Eq:qm-rel-1},~\ref{Eq:qm-rel-4},~\ref{Eq:qm-rel-8}),
can be traced back to a rotational symmetry of the 
lightcone wave functions in independent-particle models
with quarks bound by mean fields
\cite{Lorce:2011zta}. It will be interesting to
see whether the same arguments can be generalized
to antiquark TMDs.
Not all models support such relations with 
quark-target models \cite{Meissner:2007rx}
being one counter-example. However, we see that 
in the CPM the quark model relations are satisfied 
not only by quark TMDs but also by antiquark TMDs.

Let us remark that if one would impose in addition
the SU(4) spin-flavor symmetry of the nucleon wave
function, then additional relations would hold in the
CPM for antiquark TMDs the same way they hold for 
quark TMDs \cite{Efremov:2009ze,Bastami:2020rxn}.
In the quark case, the SU(4) spin-flavor symmetry
is a useful approximate concept, but it is unclear 
whether it is a useful concept for antiquark TMDs.
We therefore refrain from showing the results here,
but it will be interesting to test the SU(4)
symmetry in antiquark TMDs in future studies.

Last not least, let us remark that the Wandzura-Wilczek
(WW) approximations for the PDFs 
$g_T^{\bar q}(x)$ and $h_L^{\bar q}(x)$ 
\cite{Wandzura:1977qf,Jaffe:1991ra} are exact 
in the CPM in the antiquark case analogously to 
the quark case \cite{Zavada:2001bq,Bastami:2020rxn}.
Similarly, the similar approximate ``WW-type''
relations for the transverse moments of the TMDs 
$g_{1T}^{\perp\bar q}$ and $h_{1L}^{\perp\bar q}$
\cite{Avakian:2007mv} hold exactly in the CPM. 
Again, the proof is analogous to the quark case
\cite{Efremov:2009ze}.

\subsection{Kinematic constraints, and limitations of the approach}

The partonic interpretation of TMDs in QCD is done in infinite
momentum frame where $f_1^a(x,k_T)dx$ is interpreted as the
probability to find a parton of flavor $a$ carrying a fraction of
the longitudinal nucleon momentum in the interval $[x,\,x+dx]$
and a transverse momentum $k_T=|\vec{k_T}|$. The other twist-2 
TMDs have analogous interpretations, albeit involving polarization
of the parton and/or the nucleon \cite{Boer:1997nt}. 
Bjorken-$x$ is a Lorentz scalar, and transverse momenta 
are not affected by longitudinal boosts. Thus, we of course 
have the same $x$ and $k_T$ also in the nucleon rest 
frame (where, however, we no longer a partonic interpretation
is applicable in QCD).
If $x$ and $k_T$ are specified for an on-shell parton, then 
the parton 4-momentum $k^\mu=(E_q,\,k^1,\,\vec{k}_T)$ with
$\vec{k}_T=(k^2,\,k^3)$ is completely fixed, and we have 
\cite{Zavada:2011cv}
\bsub\label{Eq:parton-kin}\ba
    E_q &=& \frac{xM}{2}+\frac{\vec{k}_T^2+m_q^2}{2xM} 
    \label{Eq:parton-kin-Eq}\,, \\
    k^1 &=& \frac{xM}{2}-\frac{\vec{k}_T^2+m_q^2}{2xM} 
    \label{Eq:parton-kin-k1}\,.
\ea\esub

In the nucleon rest frame, the step function $\Theta_+(P\cdot k)$
introduced in (\ref{Eq:onshell-1a}) can be expressed and rewritten
thank to the on-shell condition as follows 
\be
    \Theta_+(P\cdot k) =
    \Theta(M^2+m_q^2-2ME_q)=
    \Theta(M^2-m_q^2-2M|\vec{k}|)\,.
\ee 
Hence we see that $E_q< (M^2+m_q^2)/(2M)$ and 
$|\vec{k}|\le (M^2-m_q^2)/(2M)$, i.e.\ the parton energy
and parton 3-momentum are constrained in the nucleon 
rest frame \cite{Zavada:2011cv}. 
From these bounds and the relations 
(\ref{Eq:parton-kin}) we see that $x$ and $k_T$ constrain 
each other. For instance, for a given $k_T$ the Bjorken 
variable is in the range
\bsub\be\label{Eq:bounds-x}
    \frac{1+x_{\rm min}}{2}-
    \sqrt{\Bigl(\frac{1-x_{\rm min}}{2}\Bigr)^2
    +\frac{k_T^2}{M^2}\;}
    \;\; < \;\; x \;\; < \;\;  
    \frac{1+x_{\rm min}}{2}+
    \sqrt{\Bigl(\frac{1-x_{\rm min}}{2}\Bigr)^2
    +\frac{k_T^2}{M^2}\;}
\ee
where $x_{\rm min}$ is the smallest kinematically 
possible $x$-value
\be
    x_{\rm min} = \frac{m_q^2}{M^2} \,.
\ee
The extreme $x$-values in (\ref{Eq:bounds-x}) are assumed 
only when $k_T=0$. The allowed $x$-range is then 
\be\label{Eq:bounds-x-no-kT}
    x_{\rm min}
    \;\; < \;\; x \;\; < \;\;  
    1\,.
    \ee
This is also the $x$-range in which colinear PDFs
have a finite support. In QCD, the range is $0<x<1$.
Neglecting quark masses (as it is done in all practical 
applications in QCD and in the CMP), we see that the 
model respects this constraint.
When $m_q$ is finite or $k_T$ is non-zero,
then the $x$-range is more restricted
according to (\ref{Eq:bounds-x}).
Considering a fixed value of $x$, we also
obtain a bound on the allowed transverse
parton momenta, namely \cite{Zavada:2011cv}
\be\label{Eq:bound-on-kT}
     k_T^2 \;\; < \;\;
     (1-x)\,(x-x_{\rm min})\,M^2\,.
\ee\esub

This result shows a limitation of the model. 
In QCD, a necessary condition for the applicability 
of TMD factorization is that $k_T\ll Q$. The hard scale 
$Q$ can in practice be large enough, such that transverse
momenta $k_T\sim M$ and larger may be relevant for the
phenomenological description of a DIS reaction
\cite{Schweitzer:2010tt}. However, from 
(\ref{Eq:bound-on-kT}) we see that in the CMP the
transverse parton momenta cannot exceed the bound
$k_T < \frac12\,M$. In fact, the CMP yields for the
mean transverse momenta $\la k_T\ra \sim 0.1\,$GeV
\cite{Zavada:2009ska}. The consideration 
of offshellness effects is of importance for a more 
realistic description of nonperturbative TMD properties.

\subsection{Determination of covariant functions,
and the scale of the model}

For completeness, let us briefly review the determination of the 
covariant functions ${\cal G}^a(P\cdot k)$ and ${\cal H}^a(P\cdot k)$ 
needed to obtain model predictions. The covariant functions are
Lorentz-scalars and can be evaluated in any frame. It is convenient 
to work in the nucleon rest frame where $P\cdot k=ME_q$. The parton 
energy defined in (\ref{Eq:define-Eq}) depends only on the modulus
of the parton 3-momentum. Thus,  ${\cal G}^a(P\cdot k)$ and 
${\cal H}^a(P\cdot k)$ are effectively functions of $|\vec{k}|$,
and the CMP exhibits a 3D symmetry in momentum space which tightly
connects longitudinal and transverse parton momenta\footnote{The 
    underlying symmetry is a 3D symmetry in the nucleon rest frame.
    In any other frame, longitudinal and transverse parton momenta
    are still tightly connected, but we would not call it a 3D
    symmetry.}
and gives predictive power to the approach allowing one, e.g., 
to make predictions for the $x$- and $k_T$-dependence of TMDs based
on the knowledge of the $x$-dependence of the corresponding parton
distribution functions. 

More precisely, the knowledge of the $x$-dependence of two PDFs 
(for each of the flavors $a=u,\,d,\, \bar{u},\,\bar{d},\,\dots$)
is required to determine the covariant functions in the CMP and
hence to predict all TMDs. The obvious choice are $f_1^a(x)$ and 
$g_1^a(x)$. 
From $f_1^a(x)$ one can uniquely determine ${\cal G}^a(P\cdot k)$ and
from $g_1^a(x)$ one can uniquely determine ${\cal H}^a(P\cdot k)$.
The corresponding inversion formulas have been derived in 3 
independent ways, in Refs.~\cite{Zavada:2009ska,Efremov:2009vb}
and \cite{Efremov:2010mt} as well as in \cite{DAlesio:2009cps}. 
For the light flavors $a=u,\,d,\, \bar{u},\,\bar{d}$ the parton 
masses can be neglected, and one obtains
\bsub\label{Eq:determine-cov-functions}\ba
      {\cal G}^q(P\cdot k)
      &=& - \frac{1}{\pi M^3}\,\,
      \biggl[\frac{d}{dx}\,\frac{f_1^a(x)}{x}\biggr] \\
      {\cal H}^q(P\cdot k)
      &=&  \frac{1}{\pi x^2M^3}\,\biggl[
      3g_1^a(x) + 
      2\int_x^1\frac{dy}{y}\,g_1^a(y)
      -x\,\frac{dg_1^a(x)}{dx}
      \biggr]
\ea\esub
where it is understood that $P\cdot k = \frac12xM^2$. 
Due to the kinematics constraints (\ref{Eq:parton-kin}), 
the variable $P\cdot k$ is for massless parton
$0 < P\cdot k < \frac12 M^2$ when $m_q$ is neglected 
(or $m_qM < P\cdot k < \frac12(M^2-m_q^2)$ if
we keep track of parton masses).

Numerical predictions for quark TMDs in the pure-spin 
version of the model were presented in 
\cite{Efremov:2010mt,Bastami:2020rxn}.
In the mixed-spin version of the parton model,
one needs to introduce one additional covariant function
which describes the chiral odd polarized TMDs and which can 
be determined from, e.g., transversity
\cite{DAlesio:2009cps} similarly to
(\ref{Eq:determine-cov-functions}).

In order to determine the covariant functions
in (\ref{Eq:determine-cov-functions}) it is
necessary to use $f_1^a(x)$ and $g_1^a(x)$
from a phenomenological parametrization at 
some chosen scale $\mu^2$ which must be high 
enough for the parton model concept to be
valid. But its exact value of this scale is
unknown. In \cite{Efremov:2010mt} the scale 
was chosen to be $\mu^2=4\,{\rm GeV}^2$ and
in \cite{Bastami:2020rxn} it was chosen to
be  $\mu^2=2.5\,{\rm GeV}^2$.
As different TMDs
obey different evolution equations, the 
model and the relations among TMDs
are valid only at this scale.
TMDs strictly speaking depend on two scales,
cf.\ Sec.~\ref{Subsec:def-TMD}, but one choice 
is $\zeta=\mu^2$.

\newpage

\subsection{Comparison to other approaches in literature}
\label{Subsec:discussion-other-approaches}

It is instructive to review first the spinor description 
in free theory. The spinors of a spin-$\frac12$ 
fermion or anti\-fermion of mass $m_q$ and momentum $k^\mu$ 
polarized along the spacelike vector $n^\mu$ with $k\cdot n=0$ are
customarily denoted by $u(k,n)$ and $v(k,n)$ where we suppress 
an additional index which can assume 2 values and indicates 
a spin-up or spin-down state. 
If $n^2=-1$ it is a pure-spin state, and if $-1 < n^2 < 0$ 
it is a mixed-spin state. The spinors satisfy 
\be
    (\slashed{k}-m_q)\,u(k,n)=0\, , \quad
    (\slashed{k}+m_q)\,v(k,n)=0\, .
\ee
We choose the normalization
$\bar{u}(k,n)u(k,n)=-\bar{v}(k,n)v(k,n)=2m_q$.
The polarization is revealed by acting on the spinors 
with the Pauli-Lubanski vector 
$W_\mu = -\frac12\varepsilon_{\mu\nu\rho\sigma}
\hat{J}^{\nu\rho}\hat{P}^\sigma$, where 
$\hat{J}^{\mu\nu}$ and $\hat{P}^\nu$ are
respectively the generators of rotations and translations 
of the Poincar\'e group, projected on $n^\mu$ as follows
\be
    \frac{(-W\cdot n)}{m_q}\,u(k,n)=\pm\,\frac12 \,u(k,n)\,,
    \quad
    \frac{(-W\cdot n)}{m_q}\,v(k,n)=\pm\,\frac12 \,v(k,n)\,,
\ee
with the two signs depending on whether the particles
are spin-up or spin-down. If in the particle rest frame
one chooses $n^\mu = (0,0,0,1)$, then the operator 
$(-W\cdot n)/m_q$ coincides with the $z$-component 
of the familiar intrinsic spin operator. 
The spinors satisfy the completeness relations 
\bsub\ba
    u(k,n)\otimes \bar{u}(k,n) &=& (\slashed{k}+m_q)P(n)\,
    \label{Eq:spin-dens-mat-q}\,,\\ 
    v(k,n)\otimes \bar{v}(k,n) &=& (\slashed{k}-m_q)P(n)\,
    \label{Eq:spin-dens-mat-qbar}\,.
\ea
In (\ref{Eq:spin-dens-mat-q},~\ref{Eq:spin-dens-mat-qbar})
a summation over spin-up and spin-down states is
understood (we recall that we suppress the corresponding
index, see above), and we introduced the matrix
\be
    P(n) = \frac12 (1+\gamma_5\slashed{n})
\ee\esub
For a pure-spin state, $P(n)$
is a projector, i.e.\ $P(n)^2 = P(n)$. 
Notice that $P(n)$ projects out a fermion in 
a spin-up state with respect to $n^\mu$ 
{\it but} an antifermion in a spin-down state,
cf.\ the footnotes \ref{footnote-w-qbar} and 
\ref{footnote-lambda-qbar}.
For a mixed-spin state, $P(n)$ is a spin density
matrix with the properties ${\rm tr}\,P(n) = 2$
and ${\rm tr}\,P(n)^2 < 4$. 

At this occasion it is useful to review also 
the concept of helicity defined as expectation value 
of the Pauli-Lubanski spin operator projected on
$\vec{k}/|\vec{k}|$ for a particle with 
non-zero 3-momentum \cite{Collins-book}. 
A spin-$\frac12$ particle polarized along 
(opposite to) its direction of motion has the 
helicity $+\frac12$ ($-\frac12$). Thus,
$\lambda$ in (\ref{Eq:pol-def-massless-case}) 
denotes twice the helicity of a quark, while in 
(\ref{Eq:correlator-decompose-4-bar-massless}) 
it denotes (-1)$\times$twice the helicity,
see footnote~\ref{footnote-lambda-qbar}.
Notice that longitudinal boosts do not affect
helicity with the obvious exception, for massive
particles, of boosts 
into frames where the particles move backwards
(in which case the helicity flips sign) and
boosts into particle rest frames (in which
case helicity is undefined). 

We shall not review the massless case here in 
detail, but only remark that $\lambda$ is 
a Lorentz-scalar for massless particles,
while the transverse spin is not affected
by longitudinal boost. This leads to the
description of massless quarks and antiquarks
introduced (\ref{Eq:pol-def-massless-case}) and
(\ref{Eq:correlator-decompose-4-bar-massless}) 
where $\lambda$ is twice the helicity of a
quark and $b_T^\mu$ the transverse polarization
vector for a quark (with opposite signs for
an antiquark).

This mini-review paves the way to the description
of the quark correlator $\Phi^q(k,P,S)$ in the 
free-quark-target model \cite{Tangerman:1994eh} 
where $P^\mu,\;S^\mu$ denote momentum and spin 
vector of the quark target while $k^\mu$ and $n^\mu$
are the momentum and spin vector of the parton inside
the quark target. Based on (\ref{Eq:spin-dens-mat-q}),
the quark correlator is given by
\cite{Tangerman:1994eh}
\be\label{Eq:correlator-quark-target-model}
    \Phi^q(k,P,S) 
    = u(k,n)\otimes \bar{u}(k,n)\,
    \delta^{(4)}(P-k) 
    = (\slashed{k}+m_q)
    \frac{1+\gamma_5\slashed{n}}{2}\,
    \delta^{(4)}(P-k)
\ee
The antiquark correlator 
$\Phi^{\bar q}(k,P,S)$ is zero, 
because the probability to find
an antiquark in a quark target is of 
course zero in a free theory. 
The quark target model becomes non-trivial
and then much more interesting when
gluon interactions are included and
treated with perturbative QCD methods 
\cite{Meissner:2007rx}.

From (\ref{Eq:correlator-quark-target-model}) 
one can obtain a formulation of Feynman's parton 
model \cite{Feynman:1969ej,Feynman:1973xc} by
replacing the quark target with a nucleon target.
This is customarily formulated in terms of lightcone
coordinates and sometimes referred to as the ``free
quark ensemble model'' \cite{Tangerman:1994eh}. 
Instead of the $\delta^{4}(P-k)$ in 
(\ref{Eq:correlator-quark-target-model}) 
one has some probabilities ${\cal P}_q(k)$ to find
quarks with momentum $k^\mu$ and polarization 
$s^\mu_q(k)$ inside the nucleon target with 
momentum $P^\mu$ and polarization $S^\mu$.
Now there is also a non-zero chance to find
antiquarks with the corresponding antiquark
probabilities described by ${\cal P}_{\bar q}(k)$
and a polarization vector $s^\mu_{\bar q}(k)$.
For quarks, the polarization 4-vector is given by
\be
      s^\mu_q(k) = \lambda_q\,n^\mu_q
      + s^\mu_{qT} 
\ee
where $\lambda_q$ is the lightcone helicity,
and the helicity vector $n_q^\mu$ satisfies 
$n_q^2 = -1$ and $n_q\cdot k=0$, and it is 
also $k\cdot s_{qT} = 0$. For antiquarks,
one has analogous definitions. $\lambda_q$
and $s_{qT}^\mu$ may depend on parton 
momenta. The correlators of quarks and
antiquarks can be compactly expressed as
\cite{Tangerman:1994eh} 
\ba
    \Phi^q(k,P,S) = 
    \delta(k^2-m_q^2)
    \Biggl[
    \Theta(k^+)\,{\cal P}_q(k)
    \Bigl(\slashed{k}-m_q\Bigr)
    \Bigl(1+\gamma_5\slashed{s}_q(k)\Bigr)
    +
    \Theta(-k^+)\,{\cal P}_{\bar q}(k)
    \Bigl(\slashed{k}+m_q\Bigr)
    \Bigl(1+\gamma_5\slashed{s}_{\bar q}(k)\Bigr)
    \Biggr]\,.
\ea
Except for the lightcone formulation and
different notation, this is equivalent to the
description of quark and antiquark correlators
in this work: ${\cal P}_q(k)$ is basically
equivalent to $A_3^q$ in our work,
$\lambda_q{\cal P}_q$ to $A_8^q$ and
the momentum dependence of $s_{qT}^\mu {\cal P}_q(k)$
is contained in $A_{11}^q$, and similarly for
antiquarks. However, to the best of our 
knowledge no attempt was made in the free-quark
ensemble model to consequently explore the free
equation of motion and practically determine the
covariant functions. 

The program of practically exploring the parton 
model to describe the nucleon structure was 
pioneered in 
\cite{Zavada:1996kp,Zavada:2001bq,Zavada:2002uz}.
In these works,
the starting point was a covariant description 
of the hadronic tensor and unpolarized and polarized
DIS structure functions considering exact kinematics 
of free, relativistic, on-shell partons whose 
momentum distributions are governed by covariant 
functions ${\cal G}^a(P\cdot k)$ (for unpolarized
partons) and ${\cal H}^a(P\cdot k)$ (for polarized
partons).
The polarization vector $w_q^\mu$ was constructed
starting from   
$w_q^\mu = A\,P^\mu+B\,C^\mu+C\,k^\mu$ and the
coefficients $A$, $B$, $C$ (in general functions 
of $k\cdot S$ and/or $P\cdot k$) were determined
by imposing the constraints $k\cdot w_q=0$
and $w_q^2=-1$ (which corresponds to a pure-spin
state) \cite{Zavada:2001bq}, see also
\cite{Zavada:2013ola}.
The Callan-Gross relation between unpolarized 
structure functions and the WW
relation between polarized structure functions, 
which are both approximations in QCD,
become exact in the approach 
\cite{Zavada:1996kp,Zavada:2001bq,Zavada:2002uz,
Efremov:2004tz,Zavada:2007ww}
which also generates the Cahn effect 
\cite{Zavada:2009ska} and makes specific predictions
about quark orbital angular momentum solely 
from the relativistic motion of quarks
\cite{Zavada:2013ola}.

In Ref.~\cite{Efremov:2004tz}, by exploring an
auxiliary polarized process due to the interference 
of vector and scalar currents, the approach was
used to compute a hypothetical chiral odd structure
function and the transversity PDF $h_1^q(x)$.
In \cite{Efremov:2009ze} the model was extended
to TMDs, by introducing the concept of 
``unintegrated structure functions'' which lead
to the description of twist-2 T-even TMDs 
$f_1^q(x, k_T)$, 
$g_1^q(x, k_T)$, 
$h_1^q(x, k_T)$,
$g_{1T}^{\perp q}(x, k_T)$,
$h_{1L}^{\perp q}(x, k_T)$,
$h_{1T}^{\perp q}(x, k_T)$,
and the twist-3 TMD
$g_T^q(x, k_T)$ which has a colinear
counterpart but it was unclear how to describe
other twist-3 TMDs. This was accomplished and 
a systematic description of all twist-2 and twist-3
TMDs in \cite{Bastami:2020rxn}, where the starting 
point  was the quark spinor description 
(\ref{Eq:spin-dens-mat-q}) in combination with the
covariant functions ${\cal G}^a(P\cdot k)$ and 
${\cal H}^a(P\cdot k)$.

Prior to the latter work, in Ref.~\cite{DAlesio:2009cps}
a parton model framework was developed based on the same 
concepts as in Ref.~\cite{Zavada:1996kp,Zavada:2001bq,
Zavada:2002uz,Efremov:2004tz,Zavada:2007ww,Efremov:2009ze,
Zavada:2009ska,Efremov:2009vb,Efremov:2010mt,Zavada:2011cv,
Zavada:2013ola,Zavada:2015gaa,Zavada:2019yom,Bastami:2020rxn}, 
but centered around a consequent exploration of the equation 
of motion in the quark correlator language with the puzzling
result that in ``one parton model'' the nucleon structure is
described in terms of 2 independent covariant functions
\cite{Zavada:1996kp,Zavada:2001bq,
Zavada:2002uz,Efremov:2004tz,Zavada:2007ww,Efremov:2009ze,
Zavada:2009ska,Efremov:2009vb,Efremov:2010mt,Zavada:2011cv,
Zavada:2013ola,Zavada:2015gaa,Zavada:2019yom,Bastami:2020rxn}
in ``another parton model'' 3 independent covariant functions
are necessary for that \cite{DAlesio:2009cps}.
(As explained in Sec.~\ref{Sec-1:introduction}, the
motivation of our study was to resolve the puzzle and we 
extended \cite{DAlesio:2009cps} by systematically including
quark mass effects, the antiquark correlator, and pure-spin 
vs mixed-spin parton polarization states.)

The exploration of covariant parton models dates back
earlier works
\cite{Blumlein:1996tp,Blumlein:1996vs,Blumlein:1998nv,
Jackson:1989ph,Roberts:1996ub} where structure functions
in electron-nucleon DIS or electro-weak reactions as well
as target mass corrections were studied.  An
interesting extension of 
the parton model was carried out in the quantum statistical 
approach of Refs.~\cite{Bourrely:2005kw,Bourrely:2005tp,
Bourrely:2010ng,Bourrely:2015kla,Bourrely:2018yck}
where the nucleon was treated as a gas of massless partons
(quarks, antiquarks, gluons) in a finite size volume in 
thermal equilibrium at at a common temperature. The model
exhibits in the nucleon case a total of 8 parameters which
were fixed through fits to DIS data. This number of free
parameters should be compared with the typically 
${\cal O}$(20-25) free parameters in global fits of
PDF parametrizations, and it was argued that the statistical
model provides a physically motivated and streamlined
Ansatz for global fits \cite{Bourrely:2005kw,Bourrely:2005tp,
Bourrely:2010ng,Bourrely:2015kla,Bourrely:2018yck}.
The possibility to associate partonic motion and 
mean transverse momenta $\la k_T\ra$ with temperature
was explored in \cite{Cleymans:2010aa}. 


\section{Conclusions}
\label{Sec-11:conclusions}

We have studied the description of TMDs in the
parton model. We have explored the equations
of motion to show that the quark correlator can be 
expressed in terms of a spin density matrix which must 
be treated differently in the cases of massive 
and massless quarks.
In either case, one has a choice how to describe 
the quark polarization state. 

If one chooses to work with quarks in a pure-spin 
state, the nonperturbative information content of 
the quark correlator is described in terms of two
independent amplitudes, which can be chosen to be 
$A_3^q$ describing unpolarized TMDs and $A_8^q$ 
describing polarized TMDs.
If one chooses to work with quarks in a mixed-spin 
state, the nonperturbative information is described 
in terms of three independent amplitudes, which can 
be chosen to be $A_3^q$ describing unpolarized TMDs 
and $A_8^q$ and $A_{11}^q$ describing, respectively,
polarized chiral even and chiral odd TMDs.

One central result of our work is that we
have reconciled conflicting results in 
literature regarding how many independent 
covariant functions are needed to describe
the nonperturbative information contained
in the quark correlator in the parton model, 
namely 2 vs 3 in
Refs.~\cite{Bastami:2020rxn} vs 
\cite{DAlesio:2009cps}.
In fact, there really is only one unifying 
parton model framework to which we refer as
Covariant Parton Model (CPM) where, however, 
one can choose to work with quarks in pure- or
mixed-spin states. The pure-spin state 
version of the CPM was
explored in Refs.~\cite{
Zavada:1996kp,Zavada:2001bq,Zavada:2002uz,
Efremov:2004tz,Zavada:2007ww,Efremov:2009ze,
Zavada:2009ska,Efremov:2009vb,Efremov:2010mt,Zavada:2011cv,
Zavada:2013ola,Zavada:2015gaa,Zavada:2019yom,
Bastami:2020rxn} and describes the nucleon
structure in terms of two independent amplitudes.
The mixed-spin version of the CPM  
was studied in Ref.~\cite{DAlesio:2009cps} for
massless partons and describes quark TMDs in 
terms of 3 independent amplitudes. 

The assumption of massless quarks is natural since 
in DIS processes current quark mass effects are 
suppressed by powers of $m_q/Q \ll 1$ where $Q$ 
is the hard scale of the process. In this work, 
we have considered $m_q\neq 0$ and $m_q=0$.
The results for TMDs are the same whether one
keeps $m_q\neq0$ and neglects current quark 
mass effects at the end
\cite{
Zavada:1996kp,Zavada:2001bq,Zavada:2002uz,
Efremov:2004tz,Zavada:2007ww,Efremov:2009ze,
Zavada:2009ska,Efremov:2009vb,Efremov:2010mt,Zavada:2011cv,
Zavada:2013ola,Zavada:2015gaa,Zavada:2019yom,
Bastami:2020rxn}
or works with massless quarks from the very beginning  
\cite{DAlesio:2009cps}.
But the description of the quark spin-density 
matrix differs in the two cases. 
Keeping track of current quark mass effects has
the advantage that one can use the QCD
equations-of-motion (EOM) relations to check
consistency. In the CPM, the quarks are
non-interacting and TMDs must satisfy the EOM 
relations with (pure twist-3) tilde terms 
neglected which we have shown to be the case. 
In the quark case, we have rederived 
previous model results 
\cite{
Zavada:1996kp,Zavada:2001bq,Zavada:2002uz,
Efremov:2004tz,Zavada:2007ww,Efremov:2009ze,
Zavada:2009ska,Efremov:2009vb,Efremov:2010mt,Zavada:2011cv,
Zavada:2013ola,Zavada:2015gaa,Zavada:2019yom,
Bastami:2020rxn,DAlesio:2009cps}.

Interestingly, the CPM cannot predict the sign 
of polarized chiral odd TMDs. Information from 
other nonperturbative methods (models, lattice)
is necessary to inform the CPM and choose the
physical sign for chiral odd polarized TMDs. Once the 
physical solution is chosen using one function as
input, the CPM predicts unambiguously the signs 
of all other polarized chiral odd TMDs in agreement
with other models.
The results from many quark models refer to a low
initial scale. The precise scale at which, e.g., 
PDFs should be evaluated in the CPM
is not known. But the parton model concept is valid 
at high energies, and the initial scale of the CPM 
was, e.g., chosen to be $\mu_0^2 = 4\,{\rm GeV}^2$
\cite{
Zavada:1996kp,Zavada:2001bq,Zavada:2002uz,
Efremov:2004tz,Zavada:2007ww,Efremov:2009ze,
Zavada:2009ska,Efremov:2009vb,Efremov:2010mt,Zavada:2011cv,
Zavada:2013ola,Zavada:2015gaa,Zavada:2019yom,
Bastami:2020rxn}.
The choice of the scale is part of the model.

Another important result is that we have 
extended the treatment to include antiquarks.
The quark and antiquark correlators are 
connected to each other by a field theoretical
relation which, however, we have not imposed.
Rather, we have studied the TMDs of quarks
and antiquarks independently and used the
field theoretical connection in the end of 
day to verify the theoretical consistency.
To the best of our knowledge, we have 
presented for the first time a complete
discussion of all T-even leading and
subleading antiquark TMDs in a consistent 
framework. We have shown that in the CPM 
the antiquark TMDs satisfy 
the same linear and non-linear relations as 
quark TMDs. This result might be of interest
for modelling of antiquark effects in
phenomenology. It will be interesting to use
the model for numerical predictions of
antiquark TMDs and study their impact in
phenomenology which we leave to future
studies.

The simple covariant parton model for quark 
and antiquark correlators obtained in this work
may provide the basis for the modelling of 
unintegrated parton densities and may help to 
consistently implement TMD effects in Monte Carlo 
event generators
\cite{Collins:2005uv,Collins:2007ph,
Rogers:2007ab}. 
The approach receives general support from the
fact that the WW approximation for $g_T^a(x)$ is
supported by data with an accuracy of $40\,\%$ or
better \cite{Accardi:2009au}, and according to
first phenomenological studies the WW-type
approximation for $g_{1T}^{\perp a}$ seems to work
similarly well \cite{Bhattacharya:2021twu}.
Further phenomenological applications of the
Wandzura-Wilczek-type approximation were 
practically implemented in \cite{Bastami:2018xqd}.
It will be interesting to see whether the TMDs in
nature can be better approximated in terms of a
pure-spin or mixed-spin state model.

The model results may also be of interest
to study TMDs at small transverse
momenta $k_T \ll M$ where $M$ denotes the 
nucleon mass. In the Collins-Soper-Sterman
equations governing the evolution of TMDs,
the region of small $k_T$ is correlated 
with large impact parameters $b_T$. Not 
much is known about TMDs in $b_T$-space
in the region of large $b_T \gtrsim 1\,$fm.
Here the model results could provide useful
insights which will be explored elsewhere.


When $k_T$ is not small, the model becomes less realistic. In fact, due to the absence of interactions, the partons are on mass-shell and as a consequence of that their transverse momenta are bound from above by $k_T < \frac12\,M$. It would be interesting to explore the possibility of introducing in the CPM a way to include offshellness effects and investigate whether this can lead to a more realistic modelling of $k_T$-dependencies of TMDs. 

The model also does not exhibit the initial- or final-state interactions as encoded in Wilson lines which can generate phases and give rise to T-odd TMDs like Sivers function \cite{Sivers:1989cc} or Boer-Mulders function \cite{Boer:1997nt}. The modelling of T-odd TMDs is therefore beyond the scope of the CPM. It would be very interesting to explore the possibility of introducing the necessary phases and extend the approach to the modelling of T-odd TMDs. These topics will be left to future investigations.

\ \\
{\bf Acknowledgments.}
This work is dedicated to the memory of
Anatoli Vasilievich Efremov, an outstanding
physicist, dear colleague and wise mentor 
who made numerous important contributions 
to hadronic and spin physics in his
fruitful career over more than six decades.

\ \\
This work was supported by the National Science Foundation 
under the Award No.~1812423 and Award No.\ 2111490,
by the U.S.~Department of Energy, under the contract
no.~DE-AC05-06OR23177, under which Jefferson Science Associates, LLC operates
Jefferson Lab, and within the framework of the TMD Collaboration.

\end{document}